\documentstyle[graphics,graphicx,times,psfig]{mn2e}

\newcommand {\lsim}{\mbox{$\:\stackrel{<}{_{\sim}}\:$} }

\def\colorgrey{blue}

\def\Sec#1{{Sec\-tion~\ref{s:#1}}}

\def\Eq#1{{Eq.~(\ref{e:#1})}}     
\def\Ep#1{{~\ref{e:#1}}}        

\def\EQN#1{\label{e:#1}}        
\def\Tab#1{{Table~\ref{t:#1}}}        
\def\Fig#1{{Fig.~\ref{f:#1}}}   

\def\zabs{$z_{\rm abs}$}
\def\zem{$z_{\rm em}$}
\def\apj{ApJ }
\def\apjs{ApJS }
\def\aj{AJ }
\def\apjl{ApJl }
\def\mnras{MNRAS }
\def\parn{\par\noindent}

\def\Mpc{\hbox{Mpc}}

\def\R#1{{\mathrm{#1}}}         
\def\d#1{{\R{d}{#1}}}

\def\hMpc{\mbox{$h^{-1}$Mpc}}

\def\kms{km s$^{-1}$}

\def\cm2{cm$^{-2}$}

\def\ion#1#2{#1~{\sc #2}}
\def\HI {H~{\sc i}}

\def\Lya{Lyman-$\alpha$}
\def\Lyb{Lyman-$\beta$}

\def\unit#1#2{\mbox{{\rm #1}$^{#2}$}}

\def\Obh2{\Omega_{\rm b}h^{2}}

\def\Om{\Omega_m}

\def\Ol{\Omega_{\Lambda}}

\def\rhobar{\langle{\rho}\rangle}
\def\drho{\mbox{$\rho/\rhobar$}}
\def\G12{\mbox{$\Gamma_{12}$}}

\begin{document}

\twocolumn

\title[Density structure around quasars]{The density structure around quasars from optical depth statistics\thanks{Based on observations collected at the European Southern Observatory (ESO), under the Large Programme ``The Cosmic Evolution of the IGM'' ID No. 166.A-0106 with UVES on the 8.2 m KUEYEN telescope operated at the Paranal Observatory, Chile.}}

\def\inst#1{{${}^{#1}$}}

\author[Rollinde, Srianand et al.]{ Emmanuel~Rollinde\inst{1}, Raghunathan~Srianand\inst{1}, Tom Theuns\inst{2}, Patrick~Petitjean\inst{3}, \newauthor \ Hum~Chand \inst{1}\\
$^1$ IUCAA, Post Bag 4, Ganeshkhind, Pune 411 007, India \\
$^2$ Institute for Computational Cosmology, Department of Physics,
University of Durham, South Road, Durham DH1 3LE, UK\\ \ \ \ University
of Antwerp, Campus Drie Eiken, Universiteitsplein 1, B-2610 Antwerp, Belgium\\
$^3$ Institut d'Astrophysique de Paris -- CNRS, 98bis Boulevard Arago, F-75014 Paris, France\\ \ \ \ LERMA, Observatoire de Paris, 61 Avenue de l'Observatoire,
F-75014, Paris, France\\
}
\date{Typeset \today ; Received / Accepted}
\maketitle

\begin{abstract}
We present a method for studying the proximity effect and
the density structure around redshift z=2-3 quasars. 
It is based on the probability distribution of Lyman-$\alpha$
pixel optical depths and
   its  evolution with redshift.
 We validate the method using mock spectra obtained from
hydrodynamical simulations, and then apply it to a sample of 12 bright
quasars at redshifts 2-3 observed with UVES at the VLT-UT2 Kueyen ESO
telescope. These quasars do not show signatures of associated
absorption and have a mean monochromatic luminosity of $5.4\,\times
\,10^{31}$ h$^{-2}$\, erg\,s$^{-1}$\,Hz$^{-1}$ at the Lyman limit. 
The observed
distribution of optical depth within 10~\hMpc\ from
the QSO is statistically 
different from that measured in the general
intergalactic medium at the
same redshift.
Such a change will result from the combined effects of the
increase in photoionisation rate above the mean UV-background
due to the extra ionizing photons from the quasar radiation 
(proximity effect), and the higher density of the IGM if the 
quasars reside in overdense regions (as expected from biased galaxy
formation). The first factor decreases the optical depth
whereas the second one increases the optical depth, but our measurement
cannot distinguish a high background from a low overdensity. 
An overdensity of the order of a few is required if we use 
the amplitude of the UV-background inferred from the mean Lyman-$\alpha$ 
opacity. If no overdensity is present, then we require the UV-background
 to be higher, and consistent with the existing measurements based on standard
 analysis of the proximity effect.
\end{abstract}
\begin{keywords}
{{\em  Methods}:    data analysis -   N-body simulations    -  statistical  -   
{\em Galaxies:} intergalactic medium  - haloes - structure -  quasars: absorption  lines }
\end{keywords}

\section{Introduction}
\label{s:introduction}

The hydrogen Lyman-$\alpha$ absorption lines of the \lq Lyman-$\alpha$
forest\rq\ seen in the spectra of distant quasars, are a powerful
probe of the physical conditions in the intergalactic medium (IGM) at
high redshifts ($1.8\le z\le 6$). It is believed that most of the lines
with column density, $N_{\rm HI}\lsim 10^{14}$ \unit{cm}{-2}
originate in quasi-linear density fluctuations in which the hydrogen
gas is in ionization equilibrium with a meta-galactic UV background
produced by star forming galaxies and quasars. Non-linear effects are
unimportant and therefore the properties of the Lyman-$\alpha$ forest
are described well by just three basic ingredients: quasi-linear
theory for the growth of baryonic structure, a UV radiation field, and
the temperature of the gas (Bi 1993; Muecket et al. 1996; Bi \&
Davidson 1997; Hui, Gnedin \& Zhang 1997; Weinberg 1999; Choudhury,
Srianand \& Padmanabhan 2001a; Choudhury, Padmanabhan \& Srianand
2001b; Schaye 2001; Viel et al. 2002a).  This paradigm is impressively
confirmed by full hydrodynamical simulations (Cen et al 1994; Zhang,
Anninos \& Norman 1995; Miralda-Escud\'e et al 1996; Hernquist, Katz
\& Weinberg 1996; Wadsley \& Bond 1996; Zhang et al. 1997; Theuns et
al. 1998; Machacek et al 2000; see e.g. Efstathiou, Schaye \& Theuns
2000 for a recent review).

\parn In photoionization equilibrium, the optical depth, $\tau$, is related
to the overdensity of the gas, $\Delta\equiv \rho/\langle\rho\rangle$,
by
\begin{equation}
\tau~\propto~\Delta^2 T^{-0.7}/\Gamma_{12}\propto
\Delta^{2-0.7(\gamma-1)}/\Gamma_{12}\,.
\label{e:eqn1}
\end{equation}

\parn Here, $\Gamma=\Gamma_{12}\,10^{-12}{\rm s}^{-1}$ is the hydrogen
photo-ionization rate and $T(\Delta)$ the temperature of the gas. The
associated transmission $F=\exp(-\tau)\equiv F_{\rm o}/F_{\rm c}$ is
the observed flux ($F_{\rm o}$) divided by the estimated
continuum  flux ($F_{\rm c}$).
 Photo-ionization heating and cooling by adiabatic
expansion introduce a tight relation $T=T_0\, \Delta^{\gamma-1}$ in
the low-density IGM responsible for the Lyman-$\alpha$ forest (Hui \&
Gnedin 1997; Theuns et al. 1998). The above equation has been
extensively used, especially to probe the matter clustering (Hui 1999;
Nusser \& Haehnelt 1999; Pichon et al. 2001; Viel et al. 2002b; Croft
et al. 2002; McDonald 2003; Rollinde et al. 2003).

\parn The UV-background that causes the photo-ionization is dominated by
massive stars and quasars (Haardt \& Madau 1996; Giroux \& Shapiro
1996). 
The amplitude of the corresponding photo-ionization
rate as a function of redshift, $\Gamma(z)$, and
the relative importance of the different sources, are relatively
uncertain. %
 Fardal, Giroux \& Shull (1998) have derived the 
 \HI\ and \ion{He}{ii} photoionization history 
by modelling the opacity of the IGM, using
high-resolution observations of \HI\ absorption. 
They find  $\Gamma_{12}=1-3$ at redshift $z=2-4$.
Haardt \& Madau (2001) have combined models for the emissivity of
galaxies and quasars with calculations of the absorption of UV photons
in the IGM, and estimate $\Gamma_{12}\approx 1-2$ at redshift
$z=2-3$.
More recent observations suggest that Lyman break galaxies may
dominate the UV-background at $z=3$ (Steidel et al 2001).
In simulations,  assuming 
a standard Big Bang baryon fraction,
the value of $\Gamma_{12}$ has to be between 0.3 and 2
at a redshift $z=2-3$
in order to reproduce observed Lyman-$\alpha$ forest properties,
such as the mean transmission and
the  column density distribution
(Hernquist et al. 1996; Miralda-Escud\'e et al. 1996; 
Rauch et al. 1997; Zhang et al. 1997; Choudhury et al. 2001a;
Haehnelt et al. 2001; 
McDonald \& Miralda-Escud\'e 2001, erratum 2003; 
Hui et al 2002; Tytler et al. 2004; 
Bolton et al. 2005).
\parn An independent way for estimating
$\Gamma$ is the {\em proximity effect}.
 Locally, the UV-field may be dominated by a single source, such as a
bright quasar, leading to a deficit of absorption lines sufficiently
close to the quasar. Because the amount of absorption is in general
{\em increasing} with redshift, this reversal of the trend for
redshifts close to the emission redshift of the quasar is called the
\lq inverse\rq\ or \lq proximity\rq\ effect (Carswell et al. 1982;
Murdoch et al. 1986). The strength of this effect depends on the ratio
of ionization rates from quasar and UV-background, and since the
quasar's ionization rate can be determined directly, $\Gamma_{12}$
can be inferred. This method was pioneered by Bajtlik, Duncan \&
Ostriker (1988) but more recent data have yielded a wide variety of
estimates (Lu, Wolfe \& Turnshek 1991; Kulkarni \& Fall 1993; Bechtold 1994;
Cristiani et al. 1995; Fernandez-Soto et al. 1995; Giallongo et al.
1996; Lu et al. 1996; Srianand \& Khare 1996; Cooke, Espey \& Carswell 1997; Scott
et al 2000, 2002; Liske \& Williger 2001). Scott et al. (2000) collected
estimates from the literature which vary over almost an order of
magnitude at $z=3$, i.e. $1.5\lsim\Gamma_{12}\lsim 9$.

\parn In the standard analysis of the proximity effect it is assumed that the
matter distribution is not altered by the presence of the quasar.
The only difference between the gas close to the quasar and far
away is the increased photoionization rate
in the vicinity of the QSOs. An important consequence is
that the strength of the proximity effect should correlate with the
luminosity of the quasar but such a correlation has not been
convincingly established (see Lu et al. 1991; Bechtold 1994; Srianand
\& Khare 1996; see however Liske \& Williger 2001). It is in fact likely that the quasar will be in an
overdense region. Indeed, the presence of 
\Lya\ absorption lines with
redshift \zabs\ greater than the quasar redshift \zem\ suggests possible
excess clustering of the IGM material around QSOs (Loeb \& Eisenstein
1995; Srianand \& Khare 1996). Furthermore, in hierarchical models of
galaxy formation, the super-massive black holes that are thought to
power quasars are in massive haloes 
(Magorrian et al. 1998; Marconi \&
Hunt 2003; H\"aring \& Rix 2004), which are strongly biased to
high-density regions. If the accretion rate in quasars is close to the
Eddington limit, 
then it seems plausible that the IGM density close to
the quasar is significantly higher than the mean.

\parn Recent studies of the {\em transverse} proximity effect by Croft (2004)
and Schirber, Miralda-Escud\'e \& McDonald (2004) also suggest excess absorption over that predicted by models
that assume the standard proximity effect and isotropic quasar
emission. If this is not due to an increase in density close to the
quasar, it might imply that the quasar light is strongly beamed, or
alternatively that the quasar is highly variable. Interestingly,
neither of these affects the longitudinal proximity effect
discussed in this paper.

\parn Observations of the IGM transmission close to Lyman break galaxies
(LBGs) show that the intergalactic medium contains more neutral
hydrogen than the global average at comoving scales 1 $<~$r (Mpc) $<~$5
$h^{-1}$ (Adelberger et al. 2003).  As the UV photons from the LBGs
can not alter the ionization state of the gas at such large distances,
it is most likely that the excess absorption is caused 
by the  enhanced IGM density around LBGs.
It is worth noting that various hydrodynamical simulations
have trouble reproducing this so-called galaxy proximity effect
(e.g. Kollmeier et al. 2003, Bruscoli et al. 2003, Maselli et al. 2004, Desjacques et al. 2004).
If a similar
excess of density
  around quasar host galaxies exists and is not taken into
account, then a determination of $\Gamma$ from the proximity effect
will be biased high. 

\parn In this paper, we present a new analysis of the proximity effect
of very bright quasars observed as part of the ESO-VLT Large Programme
(LP) \lq Cosmological evolution of the Inter Galactic Medium\rq\ (PI
Jacqueline Bergeron). This new method
 allows one to infer the density structure
around quasars. The method is based on the cumulative distribution
function (CDF) of pixel optical depth, $\tau$, and so avoids the Voigt
profile fitting and line counting traditionally used. Using $\tau$
instead of the transmission, $F=\exp(-\tau)$, has the great advantage
that we can take into account the strong redshift dependence
$\langle\tau\rangle\propto (1+z)^\alpha$, with $\alpha\approx 4.5$.
\parn We begin by briefly describing the data used in this paper. We
outline the procedure in \Sec{optdepth} and illustrate it using
hydrodynamical simulations in \Sec{mock}. The application to the high
signal to noise and high resolution spectra of the ESO-VLT Large
Programme is described in \Sec{largeprogramme}.  Our analysis requires
that the density be higher close to the quasar.  Results and future
prospects are discussed in \Sec{discussion}.  Throughout this paper,
we assume a flat universe with $\Om = 0.3$, $\Ol =0.7$ and $h=0.7$.

\begin{figure*}
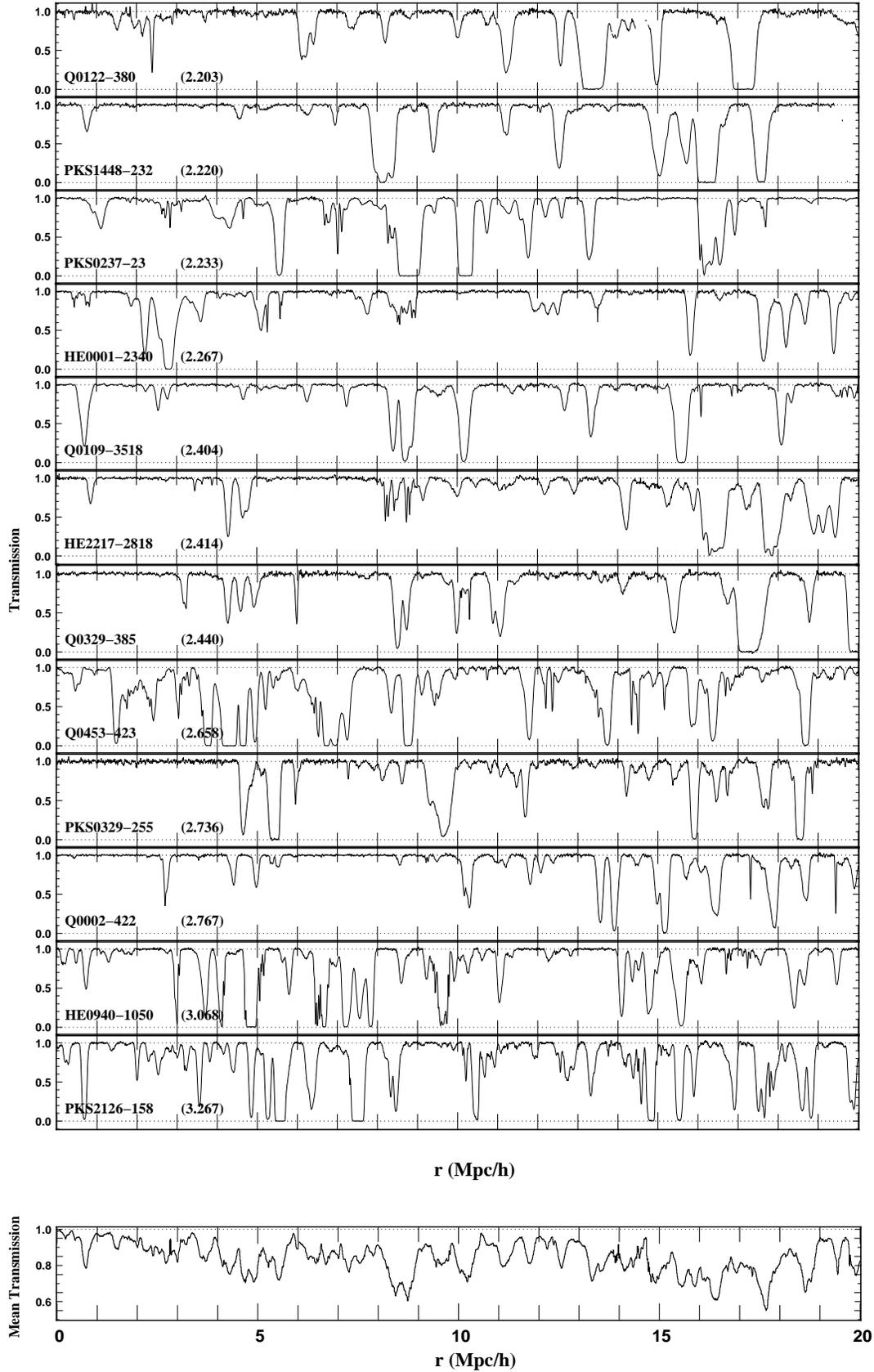

\unitlength=1cm
\begin{picture}(12,22.5)
\put(-1.43,13.19){\psfig{figure=fig1.ps,width=13.50cm,angle=-90}}
\put(-1.66,3.2){\psfig{figure=fig2.ps,width=13.73cm,angle=-90}}
\put(-1.66,0.1){\psfig{figure=fig3.ps,width=14.cm,angle=-90}}
\end{picture}
\caption{Transmission $F=\exp(-\tau)$ as a function of luminosity distance
for the LP QSOs listed in \Tab{table1}. The emission
 redshift, $z_{\rm
em}$, is indicated between brackets and increases from top to
bottom. The evolution of the optical depth with redshift 
(see \Sec{scalez}) is removed to
compute the mean transmission
$\overline{F}=\left<{\exp(-\tau/\overline{\tau}(z))}\right>$ as a
function of luminosity distance (bottom panel). The proximity effect is
clearly seen as an increase in mean transmission close to the quasar.}
\label{f:spectra}
\end{figure*}

\section{The data}
\subsection{The LP quasar sample}
\label{sect:data}
\label{s:observations}

The observational data used in our analysis were obtained with the
Ultra-Violet and Visible Echelle Spectrograph (UVES) mounted on the ESO
KUEYEN 8.2~m telescope at the Paranal observatory for the ESO-VLT Large
Programme (LP) \lq Cosmological evolution of the Inter Galactic
Medium\rq\ (PI Jacqueline Bergeron). This programme has been devised to
gather a homogeneous sample of echelle spectra of 18 QSOs, with uniform
spectral coverage, resolution and signal-to-noise ratio suitable for
studying the intergalactic medium in the redshift range
1.7$-$4.5. Spectra were obtained in service mode observations spread
over four periods (two years) covering 30 nights under good seeing
conditions ($\le0.8$ arcsec). The spectra have a signal-to-noise ratio
of $\sim$40 to 80 per pixel and a spectral resolution  $\ge 45000$ in
the Lyman-$\alpha$ forest region. Details of
the data reduction can be found in Chand et al. (2004) and Aracil et
al. (2004). In our analysis we have only used absorption lines that are
between  the \Lya\ and the \Lyb\ emission lines 
 of the quasar.

 \begin{table}
 \caption[]{ Properties of the Large Programme QSOs in our sample.
 The redshift of emission ({\em details are given in second to fourth columns})
has been  determined using different emission lines.
 The luminosity, L, in h$^{-2}$ erg\,s$^{-1}$\,Hz$^{-1}$ ({\em last column})
is computed assuming a $\Omega_m=0.3$ flat Universe and a spectral index of 0.5.}
  \begin{tabular}{|l | c c  c | c|}
  \hline
  \hline
 \multicolumn{1}{c} {quasar} & \multicolumn{3}{c}{$z_{\rm em}$}  & log(L)\\
   & mean value &  used lines & ref. &   \\               
Q0122-380    & 2.203 & H$\alpha$, \ion{Mg}{ii} & 2 & 31.633 \\                            
PKS1448-232  & 2.220 & H$\alpha$, \ion{Mg}{ii} & 2 & 31.527 \\ 		           
PKS0237-23   & 2.233 & H$\alpha$, \ion{Mg}{ii} & 2 & 31.665 \\ 		           
HE0001-2340  & 2.267 & \ion{Mg}{ii}             & 1 & 31.649 \\ 		           
Q0109-3518   & 2.404 & \ion{Mg}{ii}             & 1 & 31.819 \\ 		           
HE2217-2818  & 2.414 & \ion{Mg}{ii}             & 1 & 31.994 \\ 		           
Q0329-385    & 2.440 & H$\alpha$, \ion{Mg}{ii} & 2 & 31.278 \\ 		           
Q0453-423    & 2.658 & Lyman-$\alpha$, \ion{C}{iv}, \ion{Si}{iv} & 3 & 31.709 \\ 		           
PKS0329-255  & 2.736 & \ion{C}{iv}              & 1 & 31.577 \\ 		           
Q0002-422    & 2.767 & Lyman-$\alpha$, \ion{C}{iv}, \ion{Si}{iv} & 3 & 31.721 \\ 		           
HE0940-1050  & 3.068 & \ion{C}{iv}              & 1 & 32.146 \\ 		           
PKS2126-158  & 3.267 & Lyman-$\alpha$, \ion{C}{iv}, \ion{Si}{iv} & 4 & 32.132 \\ 	           
  \hline
  \hline
  \end{tabular} 
 \label{t:table1}
Average determinations of $z_{\rm em}$ 
are taken from Espey et al 1989 (2), 
Bechtold et al. 2002 and
Srianand \& Khare 1996 (3), using a correction
factor suggested by Fan \& Tytler 1994), Tytler \& Fan 1992 (4)
or (re)done in this paper (1, \Sec{observations}).
  \end{table}

\parn Six of the eighteen LP~QSOs \ (HE 1158$-$1843, \ HE 1347$-$2457,
\parn HE 0151$-$4326, \ \ \ \ \ HE 1341$-$1020,\ \ \ \ \ Q 0420-388\ \ \ \ \ and \parn HE 2347$-4342$) show
signatures of associated absorption close to the emission redshift of
the QSO, and are therefore excluded from our analysis. The remaining
twelve are listed in Table~\ref{t:table1}, which gives the name of the QSO, its
redshift, \zem, and the monochromatic luminosity at the Lyman limit
(L).

\begin{figure}
\unitlength=1cm
\begin{picture}(6,9)
\centerline{\psfig{figure=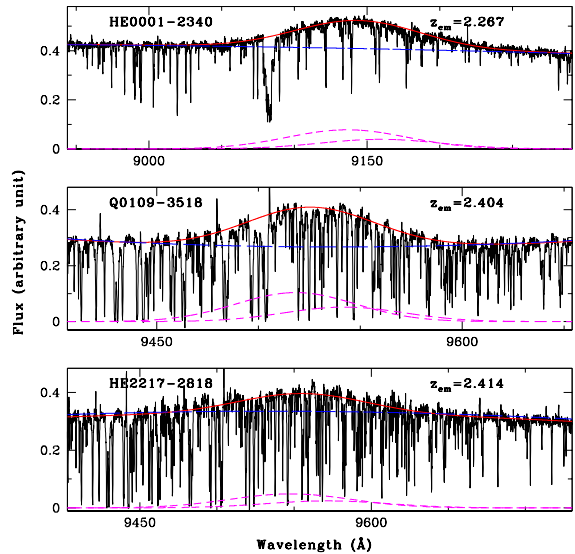,width=8cm}}
\end{picture}
\caption{Determination of the emission redshift,
\zem, 
of three QSOs using \ion{Mg}{ii} line (see \Tab{table1}).
For each quasar, the  \ion{Mg}{ii} emission lines
($\lambda\lambda$ 2796.35, 2803.53) are fitted  (two dashed lines)
on the top of a polynomial continuum (long dashed line).
The final profile is shown with a solid line. }
\label{f:fit_Mg}
\end{figure}

\parn An accurate determination of the emission redshift  is
important for the analysis.  Espey et al. (1989) have found that
the H$\alpha$ line is redshifted by an average 1000 \kms\ with respect to
lines from high ionization species and has statistically
a similar redshift as the lines from the low ionization species.  The
mean difference between H$\alpha$ and \ion{Mg}{ii} redshifts in their
sample is $\sim$ 107 \kms\ with a standard deviation of $\sim$ 500
\kms.  A redshift measurement based on H$\alpha$ and other low
ionization lines is available for 4 of the QSOs (Espey et al. 1989, see
\Tab{table1}).  We consider the mean redshift of all observed
lines for these systems.  When the \ion{Mg}{ii} emission line is
observed, as it is for three additional QSOs, we fit the profile with
the doublet of \ion{Mg}{ii} and a polynomial continuum to determine
accurately the redshift.  \Fig{fit_Mg} shows the results of this
fitting procedure for the three QSOs. On average, these
redshifts should be within an {\em rms} of 500 \kms\ from the systemic
redshift.  For 2 of the QSOs, Bechtold et al. (2002) and Srianand \&
Khare (1996) used the \ion{C}{iv}, \ion{Si}{iv} and Lyman-$\alpha$ lines to
determine the redshift of emission, and applied the correction factor
suggested by Fan \& Tytler (1994).  Otherwise, we use the \ion{C}{iv}
emission line for two other QSOs and the determination from Tytler \&
Fan (1992) for the last remaining QSO.  Therefore 7 out of 12 redshifts
of the QSOs in our sample are determined accurately using the H$\alpha$
or \ion{Mg}{ii} emission line, and 2 using the correction factor from
Fan \& Tytler (1994).
\parn 
The QSO luminosity at
the Lyman limit is computed from the available B-magnitude.
The QSO continuum slope is assumed to be a power law, $F_\lambda \sim  \lambda^{\alpha}$. We use $\alpha=-0.5$ as Francis (1993).
We checked that within a reasonable range of $\alpha=-0.5$ to $-0.7$ (e.g. Cristiani \& Vio 1990), our main result (i.e. the density structure around quasar) is not  affected by our choice of $\alpha$.
\parn 
 All possible metal lines 
and \Lya\ absorption of 
a few sub-DLA systems (there are no DLA systems
in the observed spectra) are flagged 
inside the Lyman-$\alpha$ forest.
The entire line  is removed up to the point where it 
reaches the continuum. We have not removed the
Lyman-$\alpha$ absorption associated with metal line systems
(i.e. systems with N(\HI)$<10^{19}$\unit{cm}{-2})
but the metal absorption lines
 themselves are flagged and removed.

\parn Continuum fitting of the quasar spectra is very important for our
analysis. As most of the QSOs in our sample are at lower redshifts
where line-crowding is not a problem, all the available line free
regions are used to fit the continuum. 
The procedure used to compute the continuum has
been calibrated and controlled using synthetic spectra
by Aracil et al. (2004).
They estimated that errors in the continuum
amount to about 2\%\ at   $z\sim 2.3$.
The transmission $F=\exp(-\tau)$
for each quasar in Table~1 is shown in \Fig{spectra} up to a 
luminosity distance of 20~\hMpc.

\subsection{The mock LP quasar sample}
\label{s:simulation}

We use mock spectra generated from hydrodynamical simulations to
illustrate and test the method described below. The simulated
cosmological model has $(\Omega_m,\Omega_\Lambda,h,\Omega_b
h^2,\sigma_8)=(0.3,0.7,0.65,0.019,0.9)$, where the symbols have their
usual meaning, and we have used ${\sc cmbfast}$ (Seljak \& Zaldarriaga
1996) to generate the linear power-spectrum at the starting redshift
$z=49$, assuming scale-invariant $n=1$ primordial Gaussian
fluctuations. The baryons are heated and ionized by an imposed uniform
ionizing background as computed by Haardt \& Madau (1996), and updated
by Haardt \& Madau (2001). We have increased the photo-heating rates
during hydrogen and helium reionization to satisfy the constraints on
the temperature of the intergalactic medium as determined by Schaye et
al. (2000).  This ionizing background was referred to as \lq designer
model\rq\ in that paper. In this model, hydrogen reionizes at $z=6.5$
and Helium at $z=3.5$.  The amplitude of this background is scaled so
that the mock spectra reproduce the evolution of the mean transmission
$\exp(-\tau)$ with redshift.  The simulation is performed with a
modified version of {\sc hydra} (Couchman, Thomas, \& Pearce 1995) as
described in more detail in Theuns et al (1998). {\sc hydra} combines
Smoothed Particle Hydrodynamics (SPH, Lucy 1977; Gingold \& Monaghan
1977) to represent the gas, and P3M (Couchman 1991; Hockney \& Eastwood
1981) to solve Newtonian gravity.  It follows the evolution of a
periodic, cubic region of the universe of co-moving size
$20~h^{-1}~$Mpc to a redshift $z=1.7$, using $256^3$ particles of each
species, and a co-moving gravitational softening of 20$h^{-1}$kpc.
Non-equilibrium gas cooling and photo-heating is implemented, using the
rates of Theuns et al (1998). Cold, dense gas particles are converted
to collisionless stars, but there is no feedback included. The
resolution of the simulations is close to sufficient to resolve the
Lyman-$\alpha$ forest.

\parn As the simulation is running, we store the physical state of the
IGM along many thousands of uncorrelated sight lines, which are later
patched together into mock spectra with a large redshift extent.
A full simulated spectrum typically requires around 20 individual
sightlines through the simulation box, at $z=2$. We use the
photoionization package {\sc
cloudy}\footnote{\texttt{http://www.pa.uky.edu/$\sim$gary/cloudy/}} to
compute the ionization balance of the gas in the optically thin limit,
in the presence of the Haardt \& Madau (2001) ionizing background.  We
generate 20 mock spectra for each of our observed quasars taking into
account the excess ionization by a QSO of luminosity similar to the
mean luminosity of the QSOs in our sample. A mock spectrum for a given
QSO extends over the same wavelength range as that QSO, has the same
pixel size and spectral resolution, and we add noise to the simulated
spectra with the same wavelength and flux dependence.  Except for
metals, which are flagged in the real data and are not used in this
analysis, this procedure ensures that we impose the same biases in the
reconstruction of the mock spectra, as are present in the real data.
The analysis procedure described next does not rely on simulations: we
only use simulated spectra to demonstrate that the method works.

\section{Method}
\label{s:optdepth}

\subsection{Overview}
\label{s:method}

\parn Our aim is to investigate the density structure around
high-redshift luminous quasars. We do so by investigating how the
probability distribution (PDF) of optical depths, $P(\tau)$, varies
with distance to the quasar. Far away from the QSO, $P(\tau,z)$ evolves
with redshift mainly because the mean optical depth decreases with
redshift due to the expansion of the Universe. In the appendix we show
that the {\em shape} of the PDF does not evolve much over the
relatively small redshift range $1.8\le z\le 3.1$ covered by our QSO
sample. Therefore we can define a redshift-independent {\em scaled}
optical depth distribution, $P(\tau,z)\equiv P(\tau(z)/\tau_0(z))$,
which allows us to predict the optical depth PDF at any $z$. 
The ability to take into account the strong 
redshift evolution of the mean optical depth
 is a major advantage of our method.

We can now compare this predicted optical depth PDF with the measured
one, as a function of distance $r$ to a QSO. We show that this
predicted PDF differs significantly from the measured PDF close to the
QSO. Indeed, radiation from the QSO will decrease the neutral hydrogen
fraction in its surroundings, which in turn will lead to a decrease of
the reference optical depth. This is the usual proximity effect. In
contrast if the QSO lives in a high density environment, as is
expected, then the optical depth will {\em increase}. Therefore we
need to introduce another function $f(r)$, which describes the effect
of the QSO on the PDF, such that the optical depth scales as
$\tau/(f(r)\,\tau_0(z))$. When radiation dominates, $f(r)\ll 1$, and
the optical depth becomes very small. When density dominates, $f(r)\gg
1$, and the optical depth becomes very large. The explicit expression
for $f(r)$ is given in \Eq{shiftprox} below. Of course, the
presence of the QSO might also change the {\rm shape} of the PDF. Our
main assumption in this paper is that the shape does not change, and
we demonstrate below that this is a good assumption.

By comparing the predicted to the measured optical depth PDFs, we
can determine the relative importance of radiation versus  density
enhancement. As we explain in more detail below, we can off-set a
higher amplitude of the background ionization rate with a decrease in
the over density: our determination is degenerate in this respect. So
instead of assuming no over-density and inferring the background
ionization rate, $\Gamma(z)$, as is usually done in the analysis of the
proximity effect, we will assume a given value of $\Gamma(z)$,
and recover the corresponding over-density.

This method is based on comparing optical depth PDFs. We characterise
the difference between two PDFs, by computing the maximum absolute
difference between the corresponding {\em cumulative} PDFs. Given
bootstrap re-sampled realisations of these PDFs, we can associate a
probability to a given difference in cumulative PDFs. This then allows
us to associate a given probability of the over-density as a function of
distance to the QSO, for an assumed value of the ionization rate. This
is the basis for the inferred over density as a 
function of distance to the
LP QSOs shown in Fig.~10 below.

\parn In the rest of this section we explain this procedure in more
detail, and test it on our mock QSO spectra. Readers not interested in
these details may want to skip directly to Sect.~5, where we apply the
method to the LP data.

\subsection{The optical depth - density relation}
We analyse the proximity effect using the cumulative distribution of
pixel optical depths as a function of distance to a quasar. The
starting point is \Eq{eqn1}, which relates optical depth, $\tau$, to
overdensity, $\Delta=\rho/\langle\rho\rangle$,

\begin{equation}
\tau = \tau_0\,\Delta^2 \propto \Delta^{1/(1+\beta)}\,,
\label{e:tau}
\end{equation}

where $1/(1+\beta)=2-0.7(\gamma-1)$, and

\begin{eqnarray}
\tau_0 &=& 0.206\,\left(\frac{\Omega_bh^2}{0.02}\right)^2\,
\frac{X}{0.24}\,\frac{X+0.5Y}{0.88}\,
\frac{\alpha(T)}{\alpha(T_4)}\,\nonumber\\
&\times & \frac{1}{\Gamma_{12}}\,\frac{H(z=2)}{H(z)}\,
\left(\frac{1+z}{3}\right)^6\,,
\label{e:tau0}
\end{eqnarray}

\parn is the Gunn-Peterson (Gunn \& Peterson 1965) optical depth. Here,
$\alpha(T_4=10^4K)=4.19\times 10^{-13}{\rm cm}^3\,{\rm s}^{-1}$ is the
hydrogen recombination coefficient (Verner \& Ferland 1996) which
scales approximately $\propto T^{-0.7}$ close to $T=10^4K$, $H(z)$ is
the Hubble constant at redshift $z$, $X$ and $Y$ are the hydrogen and
Helium abundances by mass, respectively, and $\Omega_bh^2$ is the
baryon fraction. We have assumed that hydrogen and helium are both
almost fully ionized. The exponent $\gamma$ and normalisation $T_0$ of
the temperature-density relation $T=T_0\Delta^{\gamma-1}$, have been
measured by e.g. Schaye et al. (2000) to be in the range
$\gamma=[1-1.5]$ and $T_0\approx 10^4$K in the redshift interval $2\le
z\le 3$. How are the density and optical depth PDFs related?\\

Let $P_\Delta(\Delta,z)d\Delta$ be the density distribution at
redshift $z$.  The probability distribution function (PDF) for the
optical depth $P_\tau(\tau,z)d\tau$ is obtained by combining
$P_\Delta(\Delta,z)d\Delta$ with \Eq{tau}. At two different redshifts
$z_1$ and $z_2$, say, $P_\tau(\tau,z) d\tau$ will differ because
$\tau_0$ changes (see \Eq{tau0}) and because the density PDF,
$P_\Delta(\Delta,z)d\Delta$, evolves as structure grows. For the
relatively small redshift range covered by the LP quasars, we show
below that the redshift evolution of $P_\tau(\tau,z)\,d\tau$ is
dominated by that of the mean optical depth, $\tau_0$, and that the
shape of the distribution does not change very much.  This is true for
the simulated quasar sample as well. The PDF of $\tau$ is therefore
given by

\begin{equation}
P_\tau(\tau,z)\,d\tau\approx
(1+\beta)\,P_{\Delta}\left[\left(\frac{\tau}{\tau_0}\right)^{1+\beta}\right]\,\left(\frac{\tau}{\tau_0}\right)^\beta\,\frac{d\tau}{\tau_0}\,,
\EQN{pscale}
\end{equation}

\parn and to a very good approximation, its redshift dependence is
through $\tau_0(z)$ only. Therefore, given the PDF of $\tau$ at
several redshifts covered by the LP sample, $1.7\le
z\le 3.1$,  one can accurately
predict the scaling factor required to scale each PDF 
$P_\tau(\tau,z)$
to the PDF observed at a given reference 
redshift, $z=2.25$.
  We will call this the {\em scaled optical depth PDF}
below.  We emphasise here that the transmission is non-linearly
related to the density. Since the median optical depth corresponds to
a value of the flux within the noise around the continuum, the
evolution with redshift cannot be taken into account with the
transmission only.

\parn Thermal broadening and peculiar velocities prevent the unique
identification of an overdensity, $\Delta$, in real space, with a
given optical depth, $\tau$, in redshift space. Therefore
$P_\Delta(\Delta,z)\,d\Delta$ does not refer to the real space over
density, but the optical depth weighted overdensity, as used for
example in Schaye et al. (1999). In the Appendix we discuss a fitting
function of $P_\Delta$ which is based on the fit introduced by
Miralda-Escud\'e, Haehnelt \& Rees (2000) for the density distribution
of the IGM. We show there that the shape of this function fits
$P_\Delta$ well, but the best fitting parameters differ considerably
from the real space density PDF. We also show that, in simulations,
$P_\Delta$ varies little with redshift in $1.7\le z \le 3.1$.

\parn A quasar's proximity effect will change the PDF of $\tau$.  The
change due to the increase in ionization rate can be accurately
predicted by the appropriate scaling of $\tau_0$. However, the density
PDF may change, as is expected for biased quasar formation, which will
modify accordingly the optical depth PDF.  In our model, the {\em
shape} of the density PDF is assumed to be unaltered, only the mean
value is changed. This is our main assumption.
Physically, this implies that feedback effects from the galaxy
hosting the QSO such as winds, infall, or excess of clustering that
may modify the density distribution itself, are neglected.
The net effect of the quasar is then a rescaling of $\tau_0$. This
scaling factor is determined as a function of distance $r$ to the
quasar, by comparing the measured PDF of $\tau$ at $r$ with the
predicted one at the same redshift.  The method is based on $\tau$,
whereas what we observe is the transmission $F=\exp(-\tau)$. We
describe how to infer $\tau$ from $F$ next.

\subsection{The optical depth distribution}
At a given redshift only part of the PDF of optical depth,
$P_\tau(\tau)d\tau$, can be recovered from the observational data.
Low values of $\tau$, $\tau\le \tau_{\rm min}$, are lost in the noise,
whereas high values of $\tau$, $\tau\ge \tau_{\rm max}$, cannot be
recovered since the \Lya\ absorption is saturated. However, we can
estimate the range $\tau_{\rm min}\le\tau\le\tau_{\rm max}$ where
$\tau$ can be accurately recovered given the noise properties of the
data.

\parn By using higher-order transitions one can accurately recover
high values of $\tau$ where \Lya\ is saturated but \Lyb\ for example
is not (Savage \& Sembach 1991; Cowie \& Songaila 1998; Rollinde,
Petitjean \& Pichon 2001; Aguirre, Schaye \& Theuns 2002; Aracil et
al. 2004).  However, here we only use the \Lya\ absorption from
normalised spectra and recover $\tau$ between $\tau_{\rm min} =
-\log(1-3\sigma) \simeq 0.1$ and $\tau_{\rm max} = -\log(3\sigma)
\simeq 2.5$, where $\sigma(\lambda)$ is the rms noise as a function of
wavelength.  Note that $\tau_{\rm min}=0.1$ is a high value compared
to the actual noise in most of the spectra.  We use this limit to be
conservative.  Since we will use the cumulative probability
distribution of $\tau$ (CPDF, in the following all probability
functions implicitly refer to $P_\tau$, unless explicitly noted), we
also keep track of the number of pixels below $\tau_{\rm min}$ and
above $\tau_{\rm max}$. The CPDF of this {\em censored} representation
of the optical depth, ${\rm CPDF}_{\rm rec}(\tau)$, is
therefore a portion of the full CPDF, CPDF($\tau)\equiv
P(\tau'<\tau)$, between $\tau_{\rm min}$ and $\tau_{\rm max}$~:

\begin{equation}
  \left\{
   \begin{array}{lcll}
       {\rm {\tiny CPDF}}_{\rm rec}(\tau) &=& 0. & \tau\,<\,\tau_{\rm min}\\
       {\rm CPDF}_{\rm rec}(\tau) &=& {\rm CPDF}(\tau)  & \tau_{\rm min}\, \le\, \tau\,\le\,\tau_{\rm max}\\
       {\rm CPDF}_{\rm rec}(\tau) &=& 1. & \tau\,>\,\tau_{\rm max} \\
   \end{array}
   \right.\,.
   \EQN{PDF_censored}
\end{equation}

\parn The values of $\tau_{\rm min}$ and $\tau_{\rm max}$ depend on
redshift because the noise level $\sigma$ does, but this dependence is
very weak for our sample. This means that when we scale two recovered
PDFs to the same reference redshift, the scaled values of $\tau_{\rm
min}$ and $\tau_{\rm max}$ will no longer be the same. For example at
lower redshift (say, $z=2$) higher overdensities $\Delta \propto
(\tau/\tau_0(z=2))^{1+\beta}$ can be recovered before the line becomes
saturated than at higher redshift ($z=3$, say) because of the
evolution of $\tau_0(z)$.  Conversely, lower over densities can be
recovered at $z=3$ than at $z=2$, before the line disappears in the
noise.  This could be exploited to increase the effective recovered
overdensity range if the evolution of $\tau_0$ was strong enough.  We
describe how we scale PDFs to a common redshift next.

\subsection{Scaling of the reference optical depth $\tau_0(z)$}
\label{s:scalez}

We show in Sections \ref{s:mock_z} and \ref{s:largeprogramme} that the
shape of the censored optical depth cumulative distribution function
in both simulations and observations, is nearly independent of
redshift. These distributions refer to regions far away from the
quasar (proper distance $\ge 50$ \Mpc/$h$) where the distribution of
$\tau$ is not modified by radiation from the QSO itself. The fact that
the shape of the PDF is conserved means that redshift evolution can be
modeled accurately by a simple redshift dependence of the reference
optical depth, $\tau_0(z)$. We find the best fitting scaling
$\tau_0(z)\propto (1+z)^\alpha$ by minimising the maximum absolute
distance between scaled optical depth CPDFs (KS distance) within
different bins in redshift.  Note that the evolution of the
number of systems within a range of 
 column densities, as
used in most previous work on the proximity effect,
 is also described
as a simple scaling.  Errors in $\tau_0(z)$ are estimated using a
bootstrap resampling of chunks of proper size 10 $\hMpc$. In the next
steps, $\tau_0(z)$ is used to scale the optical depth of each pixel to
a reference redshift of $z=2.25$.

\subsection{The proximity effect}
\label{s:proxeff}

We now consider the influence of a quasar on the optical depth
distribution in the nearby IGM, scaled to the same reference redshift
using the function $\tau_0(z)$.  We consider the effect of both the
ionizing flux emitted by the quasar and that of a modified density
distribution.

\parn Let the quasar emit ionizing photons with spectrum characterised in the
usual way as 
\begin{eqnarray}
4\pi\, J_L(\nu,r) &=& \frac{L}{4\pi r^2}\,\left(\frac{\nu}{\nu_{\rm H{\sc i}}}\right)^{-\phi}\ {\rm erg}\ \,{\rm s}^{-1}\,{\rm cm}^{-2}\,{\rm Hz}^{-1}\,.
\nonumber\\
&=& 4\pi\, J_{21}(z)\times
10^{-21}\,\left(\frac{\nu}{\nu_{\rm H{\sc i}}}\right)^{-\phi}\omega(r,z)\,.
\EQN{omega0}
\end{eqnarray}
Here, $L$ is the monochromatic luminosity of the quasar at the hydrogen
ionization threshold $\nu_{\rm H{\sc i}}$. The corresponding ionization
rate is $(12.6/(3+\phi))\,J_{21}\,\omega(r)\,10^{-12}{\rm s}^{-1}$,
when one approximates the hydrogen photo-ionization cross-section with
a power-law (Theuns et al. 1998, Table {\bf
B4}). The function $\omega$ is
\begin{eqnarray}
\omega(r,z) & = & \frac{L[{\rm h}^{-2}{\rm erg}\, \unit{s}{-1}\, \unit{Hz}{-1}]/4\pi}{4\,\pi\,(r(z)[{\rm h}^{-1}{\rm cm}])^2\, 10^{-21}\,J_{21}(z)} \nonumber\\
 &\equiv& \left(\frac{r_L(z)}{r(z)}\right)^2\,.
\EQN{omega}
\end{eqnarray}
\parn Here $r(z)$ is the luminosity distance from the quasar at
redshift $z_{\rm em}$ to the cloud at redshift $z$, at the time the
photons arrive there. For a given pixel, $r$ is computed from the
absorption wavelength of that pixel and the emission redshift of the
quasar, using the equations from Phillipps, Horleston \& White (2002)
for a $\Omega_m=0.3$ flat cosmological model 
(in the future, it would be worthwile to investigate how 
our results depend on the assumed cosmology,
as initiated for the standard proximity effect analysis
by Phillipps et al. 2002).
Note that this neglects
possible infall or outflow close to the quasar. 
 All distances are computed as a luminosity distance 
in the analysis. Yet, we may also define them as
{\em proper distance} since proper and luminosity
 distances are almost equal up to 30 \hMpc\ at the redshifts 
of interest here.
All quasars
in our sample have a similar luminosity (Table.~\ref{t:table1}), they
will then have a similar value of $r_L$ when \G12\ does not vary
strongly, as is expected (e.g. Haardt \& Madau 1996; Fardal et al. 1998).

\parn The total ionization rate $\Gamma$ in the IGM is the sum of that from the
uniform background radiation, $\Gamma^{IGM}(z)$, and from the radiation from
the quasar, $\Gamma^Q(r,z)$. The increase in $\Gamma$ will shift the
PDF of $\tau$ to smaller values, without changing its shape. Very close
to the QSO, $\Gamma(r)\propto 1/r^2$ diverges, hence according to
\Eq{tau0}, $\tau_0\rightarrow 0$, which is the usual
proximity effect.

\parn However, we argued before that the quasar is likely to be in an
overdense region, which will lead to an {\em increase} in $\tau$. We
model this by assuming that the density close to the quasar is simply a
scaled-up version of that far away from the quasar, i.e.
\begin{equation} 
P_\Delta(r,(1+\Psi(r))\Delta)\,\d{\Delta}\ \equiv\ P_\Delta(\Delta)\,\d{\Delta}\,.
\EQN{density}
\end{equation}

\parn \Eq{tau} shows that this has the effect of increasing
$\tau_0$ by a factor $(1+\Psi(r))^{1/(1+\beta)}$, shifting the PDF
 of $\tau$ at a given $r$ bin, 
to higher values without changing its shape.
To simplify the notation, we will use \drho\ to refer to 
the density structure, or enhancement, around the quasar (i.e. $1+\Psi$),
and $\Delta$ to refer to the distribution 
of density  $P_\Delta$.

\parn Note that we neglect a possible variation of temperature due to
the ionizing flux from the quasar.  Since the main modification to the
ionizing background is the larger proportion of hard photons from the
quasar, we assume that the change in temperature is not large enough
to modify the optical depth distribution in a significant way.  This
argument will not be valid if the He~{\sc ii} is not
ionized. Available observations indicate the epoch of He~{\sc ii}
reionization may be probably earlier than $z\simeq 3$ (e.g. Theuns et
al. 2002).

\parn The combined effect of a density increase and extra ionizing photons
is to shift $\tau_0$ by a factor

\begin{equation}
\tau_0 \rightarrow \tau_0\,\frac{\left(\rho(r)/\rhobar\right)^{1/(1+\beta)}}{1+(r_L/r)^2}\,.
\EQN{shiftprox}
\end{equation}

\parn 
\parn The relative importance of quasar versus UV-background ionizing
photons is characterised by $r_L(z)^2\propto L/\G12(z)$ (where $\Gamma^{IGM} = \G12\, 10^{-12}$\unit{s}{-1}). 
In the absence of any temperature enhancement the optical depth
at $r$ is globally scaled compared to the optical depth in the
intergalactic medium. As a consequence, the distribution $P(\tau)$ is
simply scaled along the abscissa toward higher values in case of an
overdensity ($\drho>1$) or lower values under the influence of the
quasar ionizing flux ($\omega>0$).  
Thus, for a given distance $r$, 
there is an intrinsic degeneracy between 
the local density structure $\rho(r)$ and the
value of \G12, combined in the above scaling factor.
Therefore, if one modifies the value of \G12,
the recovered value of  $\left(\rho(r)/\rhobar\right)^{1/(1+\beta)}$, 
is scaled by a constant value $1/\G12$ when $r\ll r_L$ 
and is independant of \G12\ when 
$r\gg r_L$. 
\parn Close to or far away from the quasar, 
 this scaling is constant, which allows the shape of the 
density enhancement to be recovered.
Then, despite the fact that the absolute value of $\rho(r)$,
when $r\ll r_L$, will depend on the value of \G12\ assumed 
in the analysis; the presence of a non-uniform density enhancement
can in principle be revealed by this method.
Conversely, if the underlying density enhancement is known 
through numerical simulations e.g., or if it is neglected
as in the standard proximity effect analysis,  
\G12\ can be recovered. 
However,  neglecting overdensities
always  implies an  overestimate of \G12, irrespective 
of the method. \\

\parn We now describe how the density structure  is
recovered and how errors are estimated.

\begin{figure}
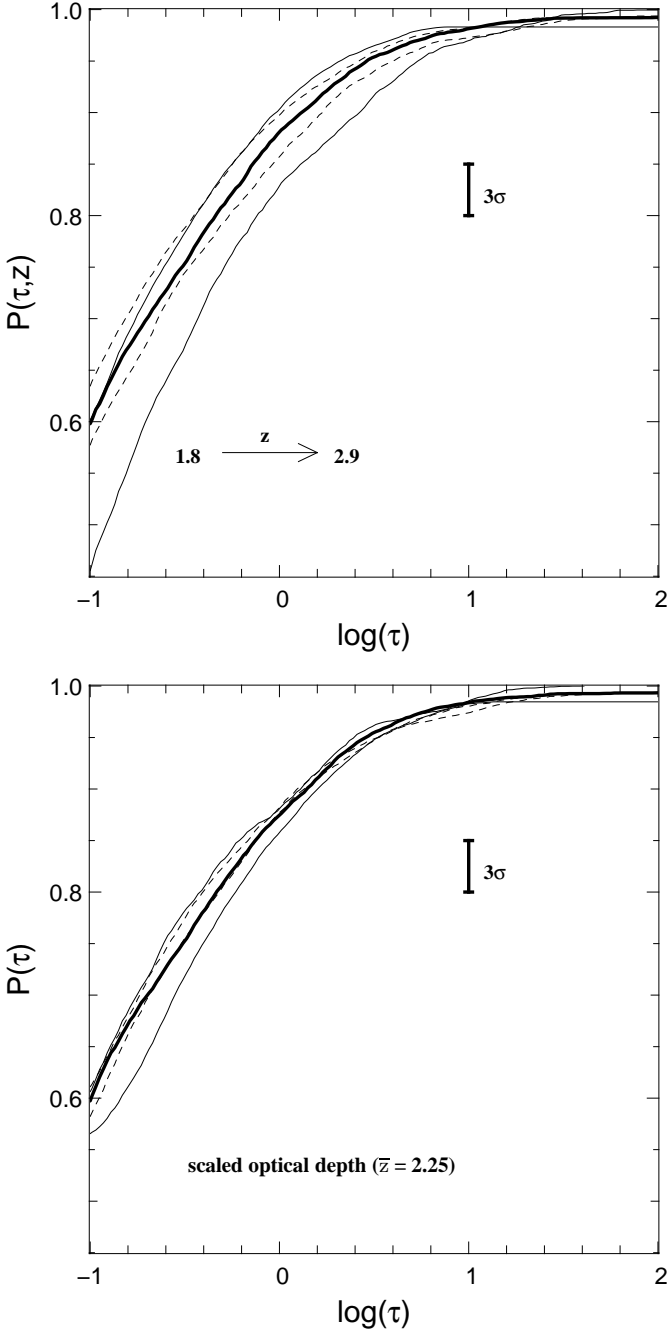

\unitlength=1cm
\begin{picture}(6,18)
\put(0,9){\psfig{figure=fig5.ps,width=\columnwidth}}
\put(0,0){\psfig{figure=fig6.ps,width=\columnwidth}}
\put(2,2){{\bf  scaled optical depth ($\overline{\rm z}$ = 2.25)}}
\end{picture}
\caption{Evolution with redshift 
of the cumulative distribution of optical depth,
computed with mock LP spectra. 
Only pixels located far  away from the proximity region
are considered.
{\em Top panel:}
 Cumulative distributions
in five bins in redshifts centred on 
$z=$1.8, 2.0, 2.25, 2.5 and 2.95 (left to right, with
alternate solid and dashed lines).
The 3$\sigma$ statistical error is 
shown with a vertical mark in both panels.
 {\em Bottom panel} Same distributions but after scaling each curve to the CPDF at
$z=2.25$ (thick curve in both panels)
by a redshift dependent scaling factor $\tau\rightarrow \tau\,\tau_0(z=2.25)/\tau_0(z)$. 
 The scaled curves are
all consistent within the 3$\sigma$ error,
 showing that the {\em shape} of the
distribution is independent of $z$. }
\label{f:cpdfvszsim}
\end{figure}

\subsection{Estimation of the density structure and errors}
\label{s:errors}		
The density structure, $\rho(r)/\rhobar$, can be inferred once the 
ionizing rate, $\G12(z)$, and the slope of the
temperature-density relation, $\gamma$, are determined. 
We will 
illustrate how \drho\ changes with changes in these parameters.

The mean scaled CPDF in the IGM, and its statistical uncertainty, are
determined from bootstrap resampling pixels outside of the possible
proximity region, at distances larger than 50 \hMpc\ proper. We
characterise the difference between two PDFs by the maximum absolute
distance (KS distance)
 between the corresponding cumulative distributions, just
as in a Kolmogorov-Smirnov test. Bootstrap resampling allows us to
associate a probability to a given value of this KS distance, ${\cal
P}(KS)$.

The proximity region is characterised by evaluating the scaled CPDF in
radial bins from the background QSO. For each radial bin, the mean
CPDF in the IGM is shifted according to \Eq{shiftprox}, using our
assumed value of $\G12$ and for different values of the function
$\left(\rho(r)/\rhobar\right)^{1/(1+\beta)}$. Given the probability associated with a given
value of KS, we can determine a probablity associated with a given
value of $\rho(r)$, $P_{\rm KS}(\rho(r)/\rhobar)$.
 The distribution of KS values of course depends on
the number of pixels in each bin. Since we want to use small bins
close to the QSO, we need to determine the probability ${\cal P}(KS)$
for each bin separately, using only pixels outside the proximity
region.

We bootstrap the QSO sample, using different sub-samples of six quasars
taken from the 12 quasars available in full sample. We can then define a
global probability associated to $\rho(r)$ as
\begin{equation} 
P(\rho(r)/\rhobar) \equiv \left < P_{\rm KS}(\rho(r)/\rhobar) \right > _{\rm sub-sample}
\label{e:prob_psi}
\end{equation}
\parn which will allow us to characterise the density structure at
different level of confidence.

\parn Note that this method is also able to recover $\G12$, if one
{\em assumes} $\rho(r)\equiv \rhobar$, i.e.  the assumption made in the
standard analysis of the proximity effect.  Indeed, the above
procedure can be done for different values of $\G12$, while maximising
the product of $P(\rho(r)~\equiv~\rhobar)$ over $r$.\\

\parn We will first apply the method to mock spectra in order to show
that this method works well.  We also use the simulations 
to show  that
our method of bootstrap sampling chunks and quasars gives realistic
errors.

\section{Proximity effect using optical depth : validation of the method with synthetic spectra}
    \label{s:mock}

\begin{figure}
\unitlength=1cm
\begin{picture}(6,9)
\centerline{\psfig{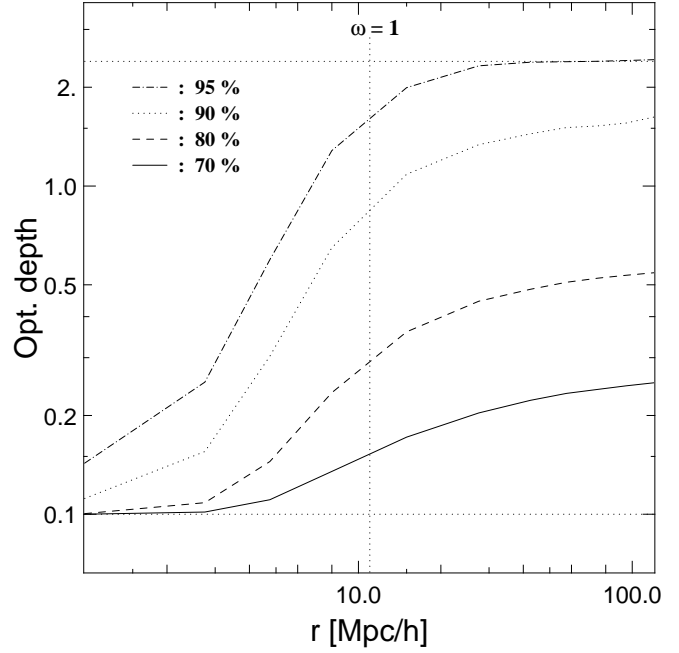}}
\end{picture}
\caption{Evolution of different percentiles of the
scaled optical depth ($z=2.25$) with luminosity
distance to the background quasar,
 in mock LP spectra. Quasars are randomly located in the 
simulation box, which implies that  
no additional density structure around
the quasars is expected statistically 
(i.e. $\drho = 1$).
Mock spectra are computed assuming 
  the mean luminosity of the LP sample, 
$L=5.4\,\times \,10^{31}$
h$^{-2}$ erg/s/Hz and \G12=1. 
The distance where the amplitude of the ionizing flux from the 
quasar and in the IGM are equal (i.e. $\omega=1$, Eq.~\Ep{omega})
is indicated by  the vertical 
dashed line.
Horizontal lines indicate
the observational upper and lower limits in  
optical depth.}
\label{f:quart_simu}
\end{figure}

\parn In this section, we use mock LP spectra, generated as described in
\Sec{simulation}. The proximity effect is implemented as described by
\Eq{shiftprox}, assuming 
the mean luminosity of the LP sample, $L =5.4\,\times \,10^{31}\
\unit{h}{-2} {\rm erg}\,\unit{s}{-1}\,\unit{Hz}{-1}$ and $\G12=1$, 
without and with additional density
enhancement. 
 Note that the value used for $\G12$ here needs not
 be equal to the value actually implemented in the simulation
itself.  
 The different steps involved in the analysis,
as described above, are now applied successively to the mock spectra.
Our assumptions and the ability of the method to recover the density
structure will be discussed.

\begin{figure}
\unitlength=1cm
\begin{picture}(6,9)
\centerline{\psfig{figure=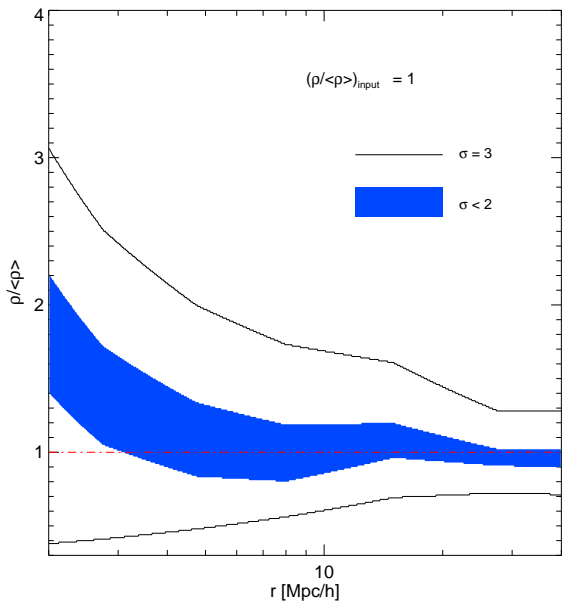,width=\columnwidth}}
\end{picture}
\caption{Recovered density structure 
 versus luminosity distance to the background quasar, from 
the analysis of mock LP spectra. Mock spectra,
including additional ionization from
the background quasar,
are generated from randomly positioned
quasar (i.e. $\rho_{\rm input}/\rhobar = 1$). 
The 2  and 3 $\sigma$
confidence levels are indicated as \colorgrey\ 
region, and solid lines respectively.
The input structure is well within the 
2$\sigma$ confidence level except for a small
bias, and a large increase of errors
below $\simeq 10$ \hMpc\ proper (luminosity
and proper distances are similar up to 30 \hMpc), 
which are explained by the modifications of the CPDF
 due to the noise (see text for details). 
The luminosity of the quasars and the 
parameters $\G12$ and $\gamma$ are identical
in the analysis and the generation of Mock spectra 
(i.e. $L =5.4\,\times \,10^{31}\
\unit{h}{-2} {\rm erg}\,\unit{s}{-1}\,\unit{Hz}{-1}$;
$\G12=1$; $\gamma=1.5$).}
\label{f:density_sim}
\end{figure}

\subsection{Evolution of the optical depth  with redshift}
\label{s:mock_z}

\parn Since we are first interested in the evolution 
of the optical depth in the IGM, we consider here
pixels at a distance larger than 50 \hMpc\ proper 
to the quasar only.  The evolution of the CPDF within
five bins in redshift centred at $z=$1.8, 2.0, 2.25, 2.5 and 2.95
is displayed in the top panel of \Fig{cpdfvszsim}.
The main evolution is driven by the mean density 
that increases, together with the mean optical depth
$\tau_0(z)$,  with redshift. This corresponds to 
a shift of the CPDF along the abscissa toward 
higher values. As explained in
\Sec{scalez}, a simple scaling of the reference optical
depth $\tau_0(z)$ is used 
to remove this primary evolution.
Parameterising $\tau_0(z)\propto (1+z)^\alpha$ gives a best fitting
 value of
$\alpha\approx 4.5$. Although  some scatter is present,
half of 50 different samples prefer a value
$4\le\alpha\le 4.5$.
Once the optical depth at each pixel is scaled using 
this relation, the CPDF computed within the same bins 
are displayed in the bottom panel of \Fig{cpdfvszsim}.
We find then that the shape is indeed conserved,
to the level of accuracy of our sample.
\parn In our mock samples, the ionizing background $\Gamma(z)$ varies
only weakly with $z$ over the range $1.7\le z\le 3.1$, as does the
temperature $T$ of the IGM.  Therefore a scaling close to $\alpha=4.5$
is indeed expected from \Eq{tau0}, given the high redshift
approximation $H(z)\propto (1+z)^{3/2}$. Below we will generate
several observed data sets by bootstrapping the LP quasars, and use
either the best fitting exponent in $\tau_0(z)\propto (1+z)^\alpha$
for each sample, or a fixed value of $\alpha=4.5$.

\subsection{Proximity effect}

\parn Once the main evolution of optical depth with redshift is
removed, we can concentrate on its change with distance to the quasar.
\Fig{quart_simu} shows the evolution of different percentiles of the
optical depth with luminosity
 distance to the mock background quasar. Note
that we only model the excess ionizating radiation from the QSO: there
is no over density at the emission redshift
(i.e. $\rho/\rhobar = 1$). 
We note that the relation between $\omega$
and distance, \Eq{omega}, depends on the luminosity of the quasar. In
our homogeneous sample, the luminosity of the QSOs, and 
then $\omega$, varies only within a factor of two
from one quasar to another. For the mock spectra, 
since we assume an unique value of the 
luminosity of each quasar,
the distance at which $\omega=1$ is 
the same for all
mock spectra: it is shown as a vertical line in the figure. 
 The effect of assuming a different luminosity  on the
recovered over density is discussed in more detail in
\Sec{largeprogramme}.

\parn \Fig{quart_simu} clearly reveals the decrease of $\tau$ with
decreasing radius, as the mock QSO starts dominating the ionization
rate. Since in this case $\drho=1$, the optical depth 
where $\omega=1$
 must be a factor of two less than its value in the ambient IGM
at $r > 50$ \hMpc\ (Eq.~\Ep{shiftprox}).  This is indeed observed
here, for each percentile.  Note how at small distances the optical
depth is everywhere decreased below $\tau_{\rm min}$, and how the
different percentiles are almost all equal to the minimum optical
depth.

\subsection{Recovery of a uniform density field}

\parn This qualitative change with distance is now studied
quantitatively to recover the underlying density field close to the
background quasars.  During the implementation of the proximity effect
in the mock spectra, we assumed $\G12=1$. Therefore, we shall use the
same value in the analysis. A wrong estimate of $\G12$ mostly leads to
a re-scaling of $\drho$ in the region of interest, close to the quasar.
Although the simulation does not correspond to a unique value of
$\gamma$ (there is a dispersion in the temperature-density relation),
the exact assumed value, if within the range specified above
(Eq.~\ref{e:tau}), does not have a large influence on the recovered
density; we assume here $\gamma=1.5$.  We will illustrate the
amplitude of these effects on the analysis of the Large Programme
quasars in Section~5.
Here, since the quasars are randomly distributed in the simulation
box, we must recover a uniform density with $\rho(r) = \rhobar$.

\parn For each bootstrap sample (\Sec{errors}), we recover a different
function $\tau_0(z)$ for the evolution of $\tau$.  However, very
similar results are obtained using a fixed evolution $(1+z)^{4.5}$,
which shows that errors on the estimation of $\tau_0(z)$ are not
essential in the analysis.  We then fit the change of the CPDF with
distance to the quasar (\Fig{quart_simu}) using \Eq{shiftprox}.  This
allows us to recover a probability distribution of $\rho(r)/\rhobar$, from the
function $P_{\rm KS}$ (Eq.~\Ep{prob_psi}).
\parn Our result is therefore expressed in terms of a probability for
each value of $\drho$ at a given radius. Different levels of
probability are shown in \Fig{density_sim}.  
The 2 and 3$\sigma$
levels of confidence correspond to the \colorgrey\
region and to the solid lines
respectively.  The input structure $\rho/\rhobar = 1$ 
 is indeed
accurately recovered at the $2\sigma$ level
for $r>$~1~\hMpc.  In this particular case,
the assumption of the standard proximity effect is satisfied (see
Introduction). Then, assuming $\drho\equiv 1$, the data (i.e. the
optical depth CPDF in our analysis, but also the mean flux
\footnote{if the distribution of $\tau$ is known between $\tau_{\rm
min}$ and $\tau_{\rm max}$, then the distribution of the flux is known
between 0 and 1, which allows us to compute the mean flux too.})  are
fitted with $\Gamma=\Gamma_{\rm true}$ within the 3$\sigma$ confidence
level. Therefore, the real value of $\G12$ may be recovered {\em if}
the density field is uniform.

\parn However, at distance lower
 than 3 \hMpc, a tendancy towards over-density
 together with a symmetric increase of errors
is apparent.
The reason is the following.
When the ionizing flux from the quasar is high
(close to the quasar), the optical depth in 
most of the pixels is below 
$\tau_{\rm min}$ (see \Fig{quart_simu}).
Then, the modeled (censored) cumulative 
function (computed from the CPDF in the IGM) 
is everywhere equal to 1.
 As for the CPDF measured directly 
in the spectra,  there will
always be a fraction of the pixels
above $\tau_{\rm min}$ due to the noise
(this fraction mostly depends on the signal to noise ratio).
Therefore, the KS distance between theoretical
 and measured CPDFs will have a maximum probability at a value larger than 0. This is not the 
case far away from the quasar, where
the theoretical CPDF, for the best fitting value of $\drho$,
 is the mean 
of all measured CPDFs.
Although most of this effect is included
in the function $P_{\rm KS}(\drho)$,
 this asymmetry will favour a value of $\drho$ higher than 1.
Besides, a lower $\drho$, that is a 
larger under-density, will not modify
the theoretical CPDF, as long as 
$\tau$ is everywhere lower than $\tau_{\rm min}$. 
This explains  the large error toward
low $\drho$ for $r\lsim 10$ \hMpc.

\begin{figure}
\unitlength=1cm
\begin{picture}(6,9)
\centerline{\psfig{figure=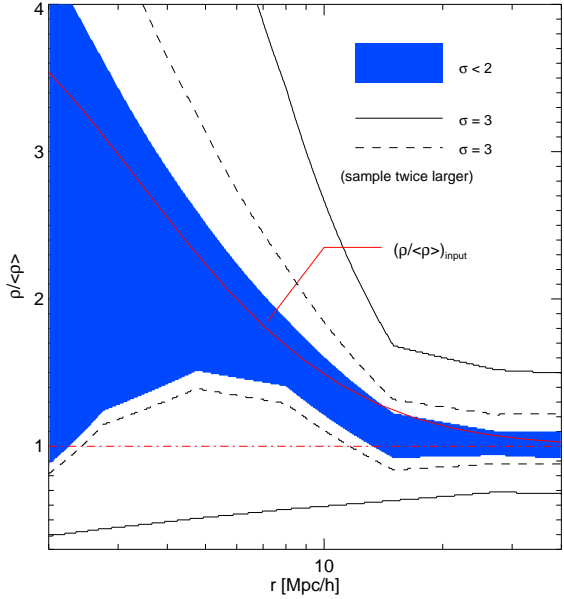,width=\columnwidth}}
\end{picture}
\caption{Recovered density structure versus luminosity
distance from 
the analysis of mock LP spectra with an additional input
density structure close to the quasar, $(\d\rho)_{\rm input}$. 
We assume $L =5.4\,\times \,10^{31}\
\unit{h}{-2} {\rm erg}\,\unit{s}{-1}\,\unit{Hz}{-1}$, $\G12=1$ and $\gamma=1.5$.
 The 2 and 3$\sigma$
confidence levels are indicated as \colorgrey\ 
region and solid lines respectively. 
The 3$\sigma$ confidence level
is also indicated for a sample twice larger 
than the large programme sample (dashed lines).}
\label{f:density_sim2}
\end{figure}

\begin{figure}
\unitlength=1cm
\begin{picture}(6,9)
\centerline{\psfig{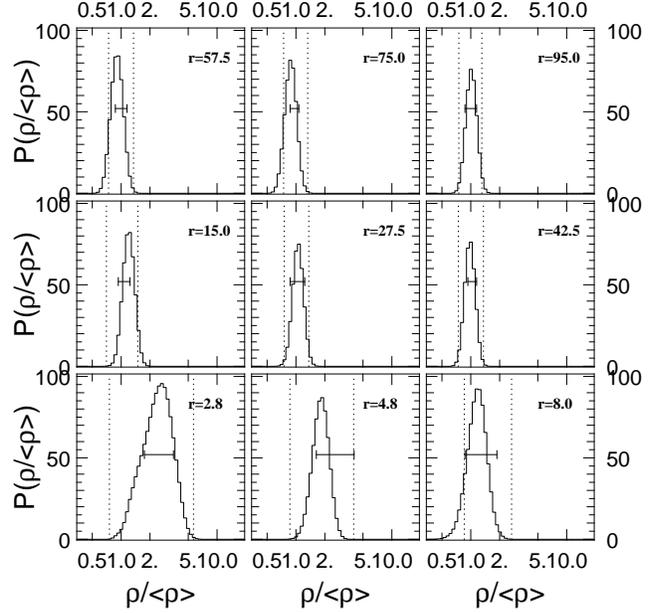}}
\end{picture}
\caption{Validation of the estimation of errors in the recovered density structure from mock spectra, with an additional
density enhancement (\Fig{density_sim2}).
 The probability distribution of $\rho/\rhobar$
obtained at different radius and with one sample is shown as a histogram in the different panels. The corresponding radius (luminosity
distance in \hMpc) is indicated and increases from left to right and bottom to top.
The range of most probable values of 
$\drho$ obtained 
from 50 different samples is indicated as an
horizontal line. 
Each estimation of the most
probable value stands between the 3$\sigma$ 
rejection level (vertical dotted lines).
}
\label{f:valid_error}
\end{figure}

\begin{figure}
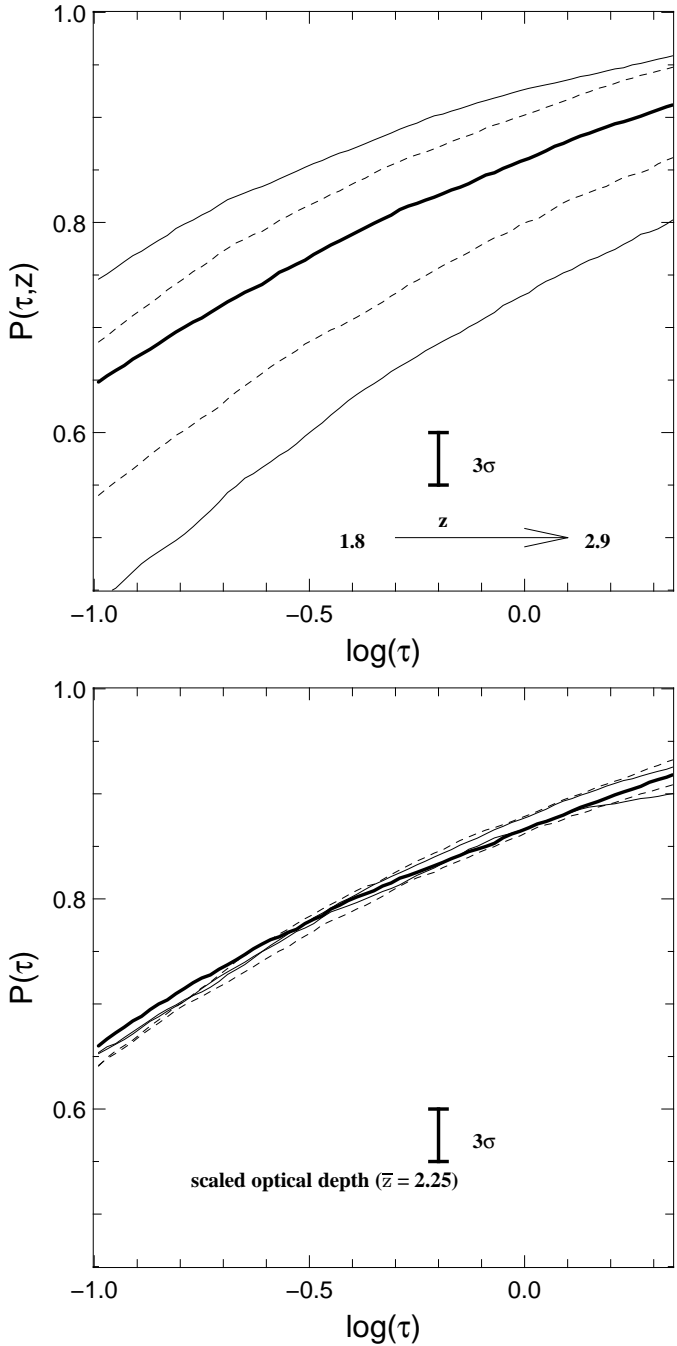

\unitlength=1cm
\begin{picture}(6,18)
\put(0,9){\psfig{figure=fig11.ps,width=\columnwidth}}
\put(0,0){\psfig{figure=fig12.ps,width=\columnwidth}}
\put(2,2){{\bf  scaled optical depth ($\overline{\rm z}$ = 2.25)}}
\end{picture}
\caption{Evolution with redshift 
of the CPDF of the optical depth, from Large
Programme spectra. Notations
are the same as in \Fig{cpdfvszsim}.
{\em Top panel:}
 Cumulative distributions
in five bins in redshifts centred on 
$z=$1.8, 2.0, 2.25, 2.5 and 2.95 (left to right, with
alternate solid and dashed lines).
 {\em Bottom panel : } Same distributions but after scaling each curve to the CPDF at
$z=2.25$ (thick curve in both panels)
by a redshift dependent scaling factor $\tau\rightarrow \tau/\tau_0(z)$.
}
\label{f:cpdfvszobs}
\end{figure}

\begin{figure}
\unitlength=1cm
\begin{picture}(6,8.5)
\centerline{\psfig{figure=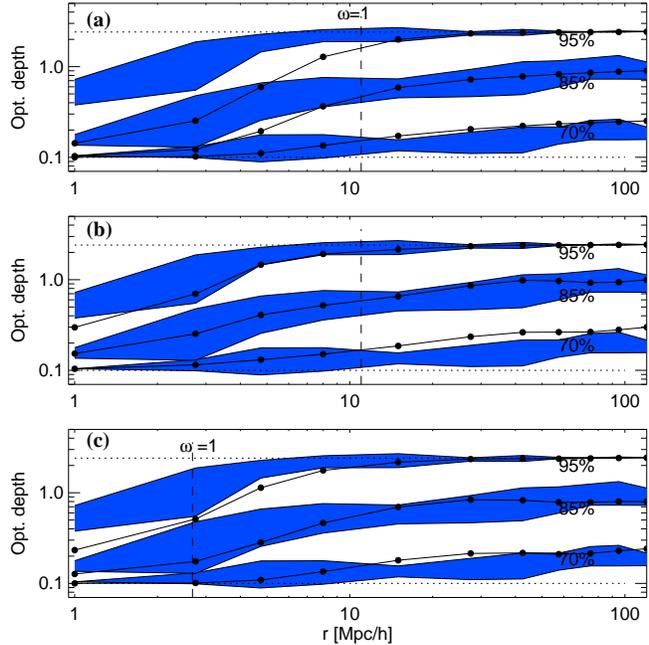,width=8.5cm}}
\put(-7.6,2.4){{\bf (c)}}
\put(-7.6,5.2){{\bf (b)}}
\put(-7.6,8.){{\bf (a)}}
\end{picture}
\caption{
Evolution of different percentiles 
(95\%, 85\% and 70\%) of the 
distribution of scaled optical depth ($z=2.25$)
with luminosity 
distance to the background quasar.
The 1$\sigma$ statistical contours corresponding to 
the Large Programme spectra (with bootstrap resampling)
are represented with \colorgrey\ regions
in each panel. 
The position where $\omega=1$ (dashed line)
is computed assuming \G12=1 in the 
panels {\bf (a)} and {\bf (b)};
and \G12=3.0 in panel {\bf (c)}.
For comparison, the mean evolution of the same percentiles
in mock LP spectra is shown 
(with solid lines and circles) assuming either \G12=1 and 
$\rho/\rhobar=1$
(panel {\bf a}, from \Fig{quart_simu});
\G12=1 and the input density structure shown in
\Fig{density_sim2} (panel {\bf b}) or
\G12=3 and $\rho/\rhobar=1$
(panel {\bf c}). A larger ionization rate or
a density enhancement are required to reproduce the observations.
Those two cases  cannot be distinguished within our 
analysis.}
\label{f:quart_LP}
\end{figure}

\begin{figure}
\unitlength=1cm
\begin{picture}(6,8.5)
\centerline{\psfig{figure=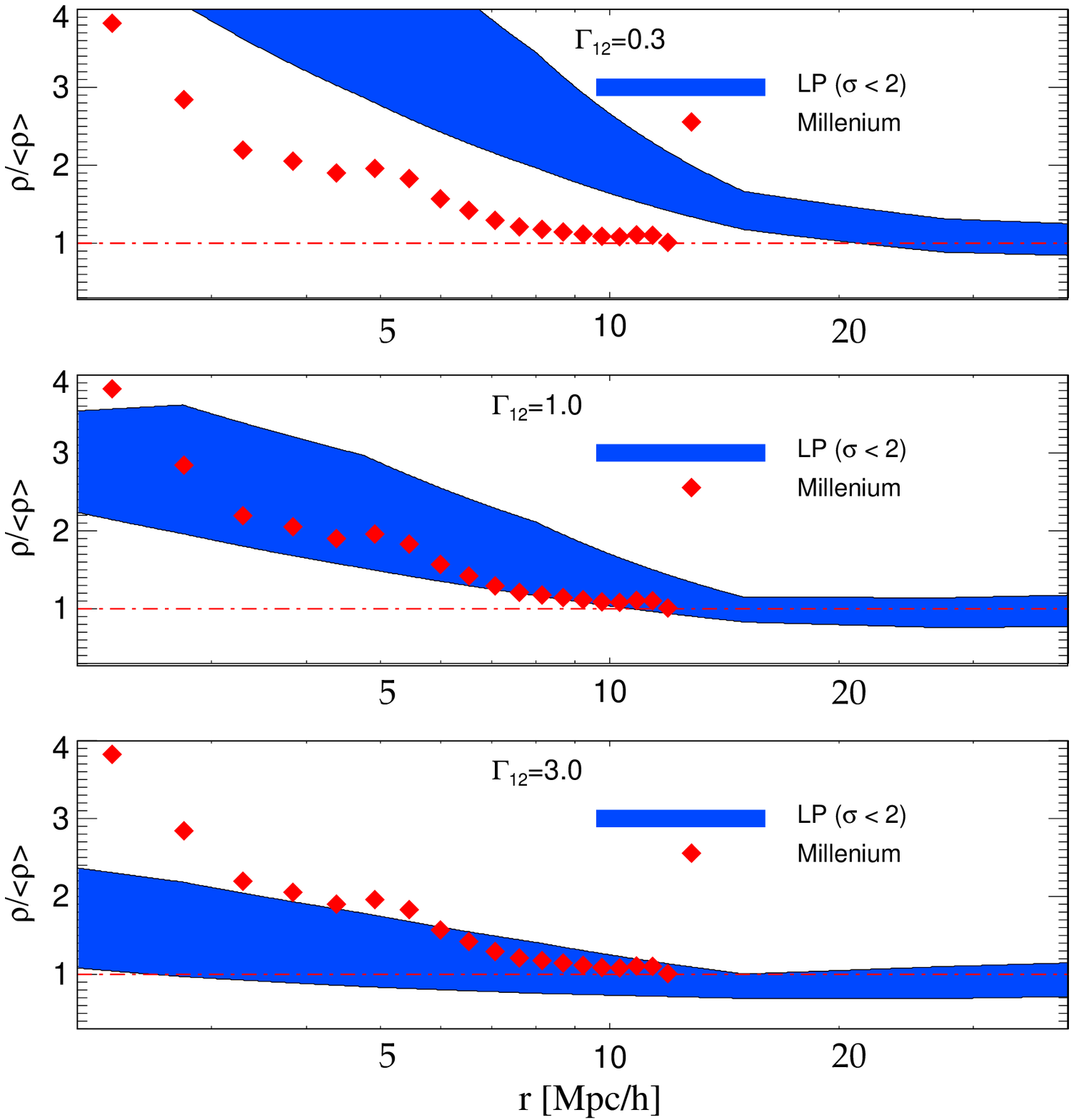,width=8.5cm}}
\put(-7.6,0.8){{\bf (c)}}
\put(-7.6,3.65){{\bf (b)}}
\put(-7.6,6.5){{\bf (a)}}
\end{picture}
\caption{
Recovered density structure 
versus luminosity distance to the background quasar from 
the analysis of the proximity effect in the Large Programme sample.
The density structure, $\rho(r)/\rhobar$,
is recovered within different bins in distance 
to the background quasar, 
from the evolution of the optical depth distribution
in the vicinity of the quasar, as compared to the 
distribution in the IGM.
Since the optical depth is a function of the density, the 
temperature and the amplitude of the ionizing flux,
the resulting density structure depends on the slope
of the temperature-density relation, $\gamma$, and on the 
amplitude of the background ionizing flux (defined by
the parameter $\G12$). The value of $\gamma$ is fixed
to 1.5 since uncertainties in it are small enough 
to have  little influence on the result.
The 2$\sigma$ contours of the recovered density
structure are shown for \G12=0.3 (panel {\bf a});
\G12=1 (panel {\bf b}) and 
\G12=3 (panel {\bf c}). 
A value of $\G12 \simeq 1$ is favored to 
recover the density enhancement derived around 
 the most massive halo at redshift $z=2$ in the Millennium
simulation  (Springel et al. 2005), which is overplotted in each panel
with diamonds.}
\label{f:density_LP}
\end{figure}

\subsection{Recovery of a density structure}
\label{s:recdens}

\parn The issue at small distances discussed above
should be less important if an overdensity is present close to the quasar.
Indeed, $\overline{\tau}$ will then remain above  
$\tau_{\rm min}$ at lower distances.
We have checked this effect by adding  
a unique density structure (directly to $\tau$, so in 
velocity space) in all spectra
with the shape $\rho(r)/\rhobar
 = 1 + 3\,\exp(-(\log(r))^2/0.6)$ (\Eq{density}).
We will show in Section~5 that, using this specific structure,
the observed evolution of optical depth percentiles is 
well fitted by the evolution in mock LP spectra (\Fig{quart_LP}).
This input structure is indicated with a 
solid line in \Fig{density_sim2}.
\parn The 2 and 3 $\sigma$ confidence levels
for the  recovered density structure 
are shown in \Fig{density_sim2}.
 It is again consistent with the 
input structure. 
As an exercise,  the analysis has been repeated
with twice as many quasars (i.e. 24). The 
corresponding contour of the 3$\sigma$ rejection level 
are  shown with dashed lines
 in \Fig{density_sim2}. The constraint
is more stringent and still in agreement
with the input structure.
As expected, the bias is not present anymore.
 Although this result
is encouraging, one must remember
that the same luminosity {\rm and}
density structure are used for all quasars, 
which would  obviously not be the case
in a real and larger sample. \\

\parn We have shown in two different cases,
with a uniform and with an enhanced density, that our analysis does recover 
the input structure. We now concentrate on the
estimation of errors.

  \subsection{Validation of error estimates}

\parn The analysis of  one sample 
(of similar properties
as the Large Programme sample), 
provides us with a probability distribution
 for the recovered density structure. 
To validate the estimation of errors,
we generate and analyse 50 different samples
of mock spectra.
In \Fig{valid_error}, the results
at different radius are reproduced
in each panel. For each radius, 
the range of most probable
values of $\drho$ obtained for each
sample is indicated by a thin horizontal line, while
a specific probability distribution 
corresponding to one sample is shown.
This procedure is done in the case of an additional
density enhancement (\Fig{density_sim2}).
The best fitting value from different realisations 
does always fall
within the 3$\sigma$ rejection level
estimated from a single sample. The same validation has
been done without additional density structure. The conclusion
is the same, although the bias discussed above implies that
 the distribution at low radius
is extended toward lower values while the best fitting
value is shifted toward higher values. \\

\parn Our analysis has been  successfully 
tested with a numerical simulation,
for the most probable result as well as the estimation of errors.
We may now turn to the analysis of the ESO-VLT Large Programme.

\section{Application to the ESO-VLT Large Programme}
\label{s:largeprogramme}

\parn 
In this section, we perform, with the LP quasars,
 the same sequence of  analysis described above. First, we have 
confirmed that the evolution of the 
mean transmission $\langle F\rangle$ 
with $z$ is consistent with previous determinations (e.g. Press, Rybicki \& Schneider 1993;
Schaye et al. 2003). In particular, this gives confidence in the continuum fitting 
procedure. 
\parn Then, we compute the evolution of optical depth with redshift, 
displayed in \Fig{cpdfvszobs} (upper panel).
It is stronger than in the mock spectra and 
seems to favor $\alpha=6$, when
fitted with $\tau_0(z)\propto (1+z)^{\alpha}$. 
However, a slope of 4.5 is allowed within
  a 3$\sigma$ confidence level.
 More important, our 
results are not modified, within the statistical errors,
whether we use $\alpha=4.5$ or the actual fit.
 The CPDF of the scaled optical depth
is shown in \Fig{cpdfvszobs} (bottom panel)
with the best fitting result for $\tau_0(z)$. Observations are also
 consistent
with the assumption that the shape of the CPDF does not
evolve from $z=3.2$ to $z=2.2$.
\parn Once the evolution with redshift is removed,
 the scaled optical
depth CPDFs are computed within different bins in distance to the
quasar. The 1$\sigma$ statistical contours (from a bootstrap
resampling) of the evolution of different percentiles 
are shown in \Fig{quart_LP} (grey regions in each panel).
 We note here that the different
percentiles are scaled roughly
by the same amount at any given radius
(when the lowest contour of $\tau$ is larger than the minimum
value, i.e. the lower dotted line).
This corresponds to the fact that 
the shape of the CPDF is conserved when one
gets closer to the quasar (at the level of accuracy of our sample).
This gives confidence in our main assumption
 that a simple scaling of the
reference optical depth is sufficient.
\parn In order to recover the density structure,
values of \G12\ and $\gamma$ have to be fixed first.
As mentioned earlier, the expected value of $\gamma$ 
is between 1 and 1.5 and we use $\gamma$ = 1.5 in most of our
analysis. The value of \G12 is between 0.3 and 3 (aside from 
measurements from standard proximity effect analysis), we use 
\G12 = 1. 
In the previous section, the mean
 evolution of optical depth percentiles was computed
in mock LP spectra 
without additional  density structure 
and using \G12 =1 (see \Fig{quart_simu}). It
is overplotted for comparison in \Fig{quart_LP}, panel
{\bf (a)}. 
In the data, there is no
clear change in the percentiles at a radius where $\omega=1$ 
(for $\G12=1$)
 and even at the lowest radii considered here, the highest
percentiles do not reach the minimum optical depth. 
In contrast, the presence of the ionizing
photons from the QSO  already strongly affects
 the optical depth percentiles in mock spectra. 
Thus, the addition of a density structure is required
to counterbalance the increase of the ionization rate.
This is shown in panel {\bf b} where 
we overplot the evolution of optical depth percentiles in mock spectra
including a density structure around 
the quasar, as described in \Sec{recdens} (\Fig{density_sim2}).
This provides then a better fit to the observed evolution. 
The probability distribution of $\drho$ associated to 
the Large Programme QSOs is directly recovered through 
the procedure described in \Sec{optdepth}. 
The 2$\sigma$ confidence region  
is then displayed in \Fig{density_LP},
again for \G12=1 and $\gamma=1.5$ (panel {\bf b}, \colorgrey\ region).
 {\em A uniform density is rejected
at the 2$\sigma$ level for $r\lsim 10$ proper \hMpc.}
\parn This recovered profile can then be compared to expected 
density  profile from simulation. For this purpose, we have used 
the Millennium simulation (Springel et al 2005).
 This dark-matter only simulation evolved $2160^3$
particles in a box of size 500 \hMpc,
 and has $\Omega_m=0.25$ and
$\sigma_8=0.9$.
Since the LP quasars are very luminous, we extract
the averaged density profile around the
 most massive halo at redshift $z=2$ in the simulation.
The profile, smoothed over 2.5 \hMpc\ is shown as diamonds
in \Fig{density_LP}. The similarity is encouraging, in particular
the fact that both profiles start to increase at the same radius 
$\simeq$  10 \hMpc. 

\parn The effect of varying $\G12$ and $\gamma$ is investigated
next.
It is reasonable to assume that $\gamma$ is within 
1 and 1.5 (see Eq.~\ref{e:tau}). Since we actually
recover $(\drho)^{2-0.7(\gamma-1)}$, varying $\gamma$
only scales \drho\ in a logarithmic plot.
The effect is negligible
compared to statistical errors.
 As for $\G12$, we have shown
in \Sec{optdepth} that, for $r\lsim r_{\rm L}$,
 $\drho$ is proportional to $(1/\G12)^{1+\beta}$.
Therefore,  
the observed optical depth percentiles evolution
could also be reproduced in mock spectra with a larger value of \G12,
which decreases the influence of the quasar ionizing flux
(the radius where $\omega=1$ is shifted towards
lower distance).
The same quality of fit in \Fig{quart_LP} (panel {\bf c}) 
is indeed obtained 
with the evolution of optical depth percentiles in mock spectra
without density structure but with a larger 
value of \G12 (3 instead of 1).
Similarly, the 2$\sigma$ confidence region
of \drho\
 is shown for \G12=3 in panel {\bf (c)}
of \Fig{density_LP}. The recovered density structure is
reduced, and  a uniform density can 
 be rejected at the 2$\sigma$ level for $r\lsim 2$\hMpc\ only.
Yet, this may as well be explained by the 
systematic bias in the recovered
structure at small distances (\Fig{density_sim}).
Higher values of \G12\ would
result in an under density at small distances.
Conversely, a lower value of $\G12$ enhances
the recovered density structure, which is demonstrated
for $\G12=0.3$ in panel {\bf (a)} of \Fig{density_LP}. 
\parn One may then ask the question of  which value of 
$\G12$ will allow the observation to be consistent with a 
uniform density $\drho=1$. This corresponds
to the standard
proximity effect applied to optical depth statistics.
If one requires that an uniform density is 
not rejected at more than 2$\sigma$, within each bin in distance,
$\G12$ is constrained to be within the range 3.6-15. This is
consistent with the range of estimates obtained from standard 
proximity effect analysis 
using  line counting statistics ($\G12\simeq 1.5-9$). 
We could also assume the density profile
based on the  Millenium simulation to recover \G12.
In this case, \Fig{density_LP} shows that 0.3$< \G12\ \lsim 3$.

\section{Conclusion}                 
\label{s:discussion}
\label{s:conclusion}

\parn In this article we presented a method to
probe the density structure around quasars,
using a new analysis of  the proximity effect 
in absorption spectra of quasars. 
In the vicinity of the quasar, the additional ionizing photons 
increase the total ionizing rate which decreases 
the Lyman-$\alpha$ absorption. Simultaneously, an
increase of the density around the quasar (as expected from biased galaxy
formation) would increase the absorption. Both effects are
better probed with the optical depth than directly with the flux.
Our method also avoids fitting the individual absorption
lines, and directly uses the  cumulative distribution
 of Lyman-$\alpha$ optical depths observed in each pixel.
We then  model
the change of this distribution under modification
of the  density field 
and the amplitude of the ionizing rate, $\G12$. 
Our method therefore allows one,
in principle, to estimate
the density enhancement around host galaxy of quasars, once 
 $\G12$ is fixed by some other method. 
\parn We first use a LCDM high resolution simulation to
 validate our method.
The information on $\G12$ and density field 
is accurately  recovered.
 This gives us confidence to perform our analysis on
the real data.
 We then use the  spectra of 12  quasars with 
highest luminosity 
at $2.2<z<3.3$ from the ESO-VLT Large Programme.\\
\parn Our method has revealed the presence of 
an overdensity for $2\lsim r \lsim 10$ proper \hMpc,
assuming $\G12<3$. 
We have shown that it is consistent with a density profile
around the most massive halo at redshift $z=2$ in the Millenium
simulation for $\G12=1$ (\Fig{density_LP}). 
In the future, a similar analysis should be done with 
a larger sample of spectra, covering different redshift 
and luminosity ranges. Together with synthetic density profiles computed
 around halos of different mass in a large simulation such as 
the Millenium one,
this will be very useful to understand better the relation 
between the environement of the quasar and its host galaxy,
and their evolution with redshift.
New constraints could also be put on the mass-luminosity relation.
\parn Without the knowledge of \G12, and due to the limited statistics,
we could not discard an uniform density profile.
Indeed, consistently with standard
proximity effect analysis, observations are also modelled
without density enhancement, assuming a higher value of \G12.
Yet, due to the specific scaling of the density profile with \G12,
a larger statistics could already allow us to distinguish
between different
type of profiles, from a simple power law to the existence
of alternate shells corresponding to
 over and under density regions. This would be valuable
to test the presence of winds, or other specific 
feedback effects. 
Thus it is important to confirm our tentative
finding of density enhancement 
around QSOs (for $\G12<3$) at high significant level using
a bigger sample.\\

\parn
Another application of this analysis concerns the 
transverse proximity effect. The modeling  of the observations
obtained with Lyman break galaxies or quasars
has been done either with simulations (Croft 2004;  Maselli et al. 2004)
or analytical model for the density (Schirber et al. 2004). 
These works could not reproduce the amplitude of
 the observed effect with normal
properties of the quasar, such as anisotropy of the beaming
and variability.
Combining the constraints on the optical depth
evolution along and transverse to 
the line of sight could be a way to disentangle 
the different parameters, that is the density
structure, \G12\ and the properties
of the quasar.

\vskip 0.5cm
\noindent{\sl Acknowledgements:} \parn We thank B. Aracil for the
reduction and the continuum fitting of the data,
and  J. Miralda-Escud\'e for useful comments.
\parn We thank
E. Thi\'ebaut, and D. Munro for freely distributing \ \ his \ \ Yorick
programming language \ \ (available \ \ at 
 {\em \tt
ftp://ftp-icf.llnl.gov:/pub/Yorick}), which~we used to implement our
analysis. 
 \parn ER was supported by a grant LAVOISIER from the French
foreign office.  HC thanks CSIR, INDIA for the grant award
No. 9/545(18)/2KI/EMR-I and CNRS/IAP for the hospitality.  RS and PPJ
gratefully acknowledge support from the Indo-French Centre for the
Promotion of Advanced Research (Centre Franco-Indien pour la Promotion
de la Recherche Avanc\'ee) under contract No. 3004-3.  
TT and PPJ acknowledge support
from the 
European RTN  program "The Physics of the Intergalactic  Medium".
TT thanks PPARC
for the award of an Advanced Fellowship, and the Kavli institute at the
University of Santa Barbara for hospitality.  
TT thanks IUCAA for hospitality
and the Royal Society for the  travel
 grant to India.
Research was conducted in
cooperation with Silicon Graphics/Cray Research utilising the Origin
super computer at DAMTP, Cambridge. This research was supported in part
by the National Science Foundation under Grant No.~PHY99-07949.

\vskip 0.5cm

%

\section*{Appendix}
\begin{figure}
\unitlength=1cm
\begin{picture}(6,8.5)
\centerline{\psfig{figure=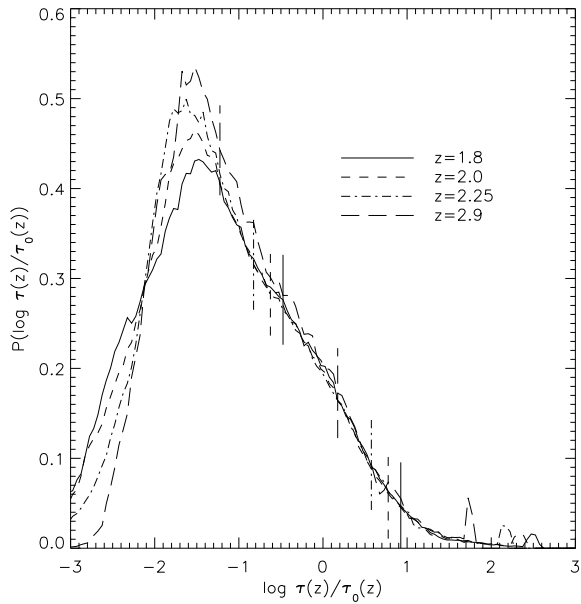,width=\columnwidth}}
\end{picture}
\caption{PDF of the true, scaled optical depth, $\tau/\tau_0$, of a
large sample (20) of mock LP quasars, in the redshift range [1.7,1.9],
[1.9,2.1], [2.15, 2.35], [2.85, 3.05]. A redshift scaling
$\tau_0\propto (1+z)^5$ is assumed pixel by pixel, the mean 
redshift is indicated in the panel. Limits in optical depth for the
censored PDFs, are indicated by thin vertical lines
(with corresponding types). The PDFs have a
Gaussian shape, with a more extended power-law tail toward low as well
as higher optical depths. The shape of the scaled PDF is almost
independent of redshift over nearly three decades in $-1\le \log(\tau/\tau_0)\le 2$.}
\label{f:pdf_fitted}
\end{figure}

A key assumption in this paper is that the PDF of the {\em scaled}
optical depth, $P(\tau(z)/\tau_0(z))$, varies little with
redshift. Here, $\tau_0(z)\propto (1+z)^{\alpha}$ is a redshift
dependent scaling function, with $\alpha\simeq 4-5$. We showed in
\Fig{cpdfvszsim} that this is true for the 
full optical depth
in mock spectra in the range $0.1\le \tau/\tau_0\le 100$ and in
\Fig{cpdfvszobs} for the censored, recovered optical depth in
the range $0.1\le \tau/\tau_0\le 2.5$, both at the reference redshift
$z=2.25$. We illustrate in \Fig{pdf_fitted} the limitation
of this assumption, by showing the scaled $P(\tau/\tau_0)$ over a
larger range. As expected, the PDF becomes wider in its tails as the
density field becomes increasingly non-linear at lower
redshifts. However in the range in which we use the PDF, $\tau_{\rm
min}\le \tau\le\tau_{\rm max}$, this dependence is very weak indeed.
It also becomes clear from this figure that we cannot reliably
determine the shape of the PDF around the maximum for the signal-to-
noise ratio in the LP quasars, even at the higher redshifts $z\sim 3$.
This is also clear from \Eq{tau}: $\tau\sim 0.07< \tau_{\rm
min}$ at the typical volume-averaged overdensity $\Delta=1/3$ at $z=3$,
when $\tau_0\simeq 0.7$. Uncertainties associated with continuum
fitting make this part of the PDF uncertain, in addition to these
signal-to-noise issues. Note that in our previous analysis we used the
recovered optical depth from mock samples, which were continuum-fitted
to mimic observed samples. This will strongly affect the shape of the
PDF at these low values, and therefore it is not very worthwhile to try
to take these lower optical depths into account for the present
analysis. In contrast, the {\em mock} PDF is uncertain at high $\tau$,
where it becomes sensitive to lack of self-shielding and other
numerical uncertainties in high density regions. Given these
limitations, can we understand the shape of the optical depth PDF in
the intermediate regime?

\parn Miralda-Escud\'e, Haehnelt and Rees (2000) provide physical motivation
for the following fitting function for the (volume-weighted, real space)
overdensity $\Delta$,

\begin{equation}
P(\Delta)\,d\Delta = A\,\exp\big[-{(\Delta^{-2/3}-C_0)^2\over 2
(2\delta_0/3)^2}\big]\,\Delta^{-\beta}\,d\Delta\,.
\end{equation}
Their Table~1 provides values for $A$, $C_0$, $\delta_0$ and $\beta$ at
redshifts $z=2$, 3, 4 and 6, which they obtained from fitting their
numerical simulations. The exponent guarantees that the PDF is a
Gaussian in $\Delta-1$ when $C_0=1$ and the dispersion $\delta_0\ll
1$.

\parn We can use this as an {\em Ansatz} for the PDF of $\tau$, given the
relation \Eq{tau} between $\Delta$ and $\tau$. We expect the
exponent in the exponential to change $-2/3\rightarrow -2(1+\beta)/3$,
and $1+\beta=[0.5,0.6]$ for $\gamma=[1,1.6]$, hence we fit
\begin{equation}
P(x)\,dx = A\,\exp\left[-{(x^{-2\nu/3}-C_0)^2\over 2
(2\delta_0/3)^2}\right]\,(10^x)^{-\mu}\,dx\,,
\label{e:fits}
\end{equation}

\begin{figure}
\unitlength=1cm
\begin{picture}(6,8.)
\centerline{\psfig{figure=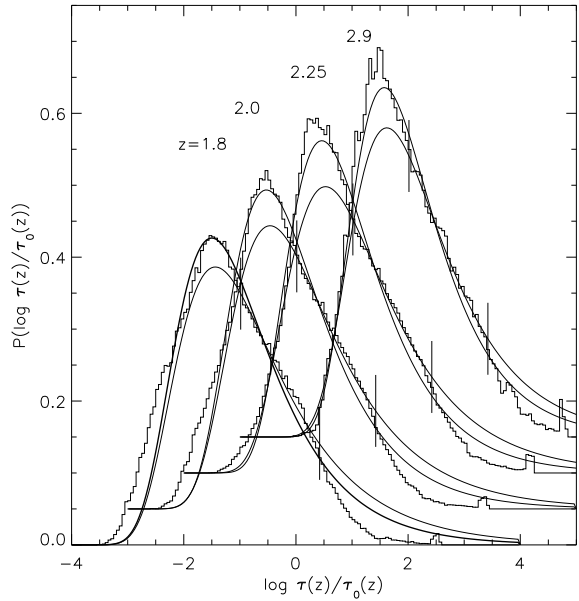,width=\columnwidth}}
\end{picture}
\caption{Fits of the form \Eq{fits} (full lines) to the
scaled PDFs shown in \Fig{pdf_fitted}, represented here by the
histograms. The fits shown by the thin line restrict the fitted region
to that of the censored optical depth (vertical lines). Different
redshift range indicated in the panel are off-set vertically and
horizontally by 0.05 and 0.1 respectively, for clarity. The fitting
function does reasonably well around the maximum and in the power-law
tail toward higher $\tau$, but is not able to fit the more non-linear
parts at very high and very low $\tau$. The fit to the censored optical
depth (thin lines) does not recover well the PDF around the maximum.}
\label{f:pdf_sim}
\end{figure}

\parn where $x\equiv \log(\tau/\tau_0)$, with free parameters $\nu\approx
1+\beta$, $C_0$, $\delta_0$ and $\mu$, and $A$ a normalisation
constant. Restricting the fit to $-2\le x\le 1$, we show the 
best fitting PDFs
in \Fig{pdf_sim} and provide the best fitting
 parameters in
Table \ref{table:fit}. The best fitting
 value for $C_0\approx 0$ is kept
constant.  The dispersion $\delta_0$ differs significantly from the
best fitting one
 to the density PDF, but the value of the exponent $\nu$ is
close to expected.

\begin{figure}
\unitlength=1cm
\begin{picture}(6,9)
\centerline{\psfig{figure=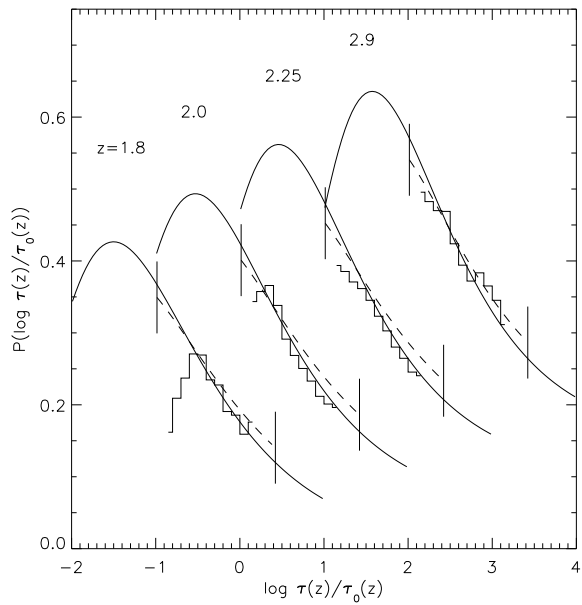,width=\columnwidth}}
\end{picture}
\caption{Overlay of the fits to the scaled PDF of the mock sample from
\Fig{pdf_fitted} to the (censored) scaled PDF of the LP quasar
sample. The same redshift scaling
$\tau_0\propto (1+z)^5$ is assumed for the LP data. 
The same redshift range are indicated and shifted as in \Fig{pdf_sim}.
 The agreement is very good, increasing our confidence that the
mock samples are sufficiently realistic for validating our method.}
\label{f:pdf_obs}
\end{figure}

\parn These fits are overlaid on the censored PDF of the observed LP quasar
sample in \Fig{pdf_obs}. The good agreement suggest that the
mock sample is indeed representative of the observed distribution.

\begin{table}
\caption[]{best fitting
 parameters (Eq.~\Ep{fits}) for the  PDF of 
scaled optical depth, within different redshift bins,
restricting the fit to $-2\le \log(\tau/\tau_0) \le 1$ (thin lines
in \Fig{pdf_sim})}
\begin{tabular}{|l | c c c }
\hline
\hline
z & $\delta_0$ & $\mu$ & $\nu$ \\
\hline
1.8   & 3.58 & 1.44 & 0.46 \\
2.0   & 3.80 & 1.46 & 0.47 \\
2.25  & 4.09 & 1.47 & 0.50 \\
2.9   & 3.80 & 1.49 & 0.52 \\
\hline
\end{tabular} 
\label{table:fit}
\end{table}

\end{document}